\documentclass[11pt]{article}
\def\etal{{\it et\thinspace al.}\ }
\def\eion{{(e~+~ion)}\ }
\def\en{{$n$}\ }
\def\no{{n$_o$}\ }
\def\art{{$\alpha_R(T)$}\ }
\def\gamr{\mit\Gamma\sp{\rm r}}
\def\gama{\mit\Gamma\sp{\rm a}}
\def\Gamr{\mit\Gamma\sp{\rm r}}
\def\Gama{\mit\Gamma\sp{\rm a}}

\usepackage{psfig}

\begin{document}


 \title{Self-Consistent R-matrix Approach To Photoionization And Unified 
Electron-Ion Recombination}
 \author{Sultana N. Nahar and Anil K. Pradhan \\
Dept of Astronomy, The Ohio State University, Columbus, OH 43210, USA}
\maketitle

\begin{abstract}
 A unified scheme using the R-matrix method has been developed for 
electron-ion recombination subsuming heretofore separate treatments of 
radiative and dielectronic recombination (RR and DR). The ab initio approach 
within the coupled channel approximation has several inherent advantages in 
addition to the natural unification of resonant and non-resonant phenomena. 
It enables a general and self-consistent treatment of photoionization and 
electron-ion recombination employing idential wavefunction expansion. Detailed 
balance takes account of interference effects due to resonances in cross 
sections, calculated explicitly for a large number of recombined \eion bound 
levels over extended energy regions. The theory of DR by Bell and Seaton
is adapted for high-$n$ resonances in the region below series limits.
The R-matrix method is employed for (A) partial and total photoionization and 
photorecombination cross sections of \eion bound levels, and (B) DR and 
\eion scattering cross sections. Relativistic effects and fine structure are 
considered in the Breit-Pauli approximation. Effects such 
as radiation damping may be taken into account where necessary. Unfiied 
recombination cross sections are in excellent agreement with measurements on 
ion storage rings to about 10-20\%. In addition to high accuracy, the 
strengths of the method are: (I) both total and level-specific cross sections 
and rate coefficients are obtained, and (II) a single \eion recombination rate 
coefficient for any given atom or ion is obtained over the entire temperature 
range of practical importance in laboratory and astrophysical plasmas,
(III) self-consistent results are obtained for the inverse processes of
photoionization and recombination; comprehensive datasets have been computed 
for over 50 atoms and ions. Selected data are presented for iron ions.

\end{abstract}

keywords(Photoionization, Electron-ion Recombination)


\section{Introduction}
\label{}

 The electron-ion recombination process is unified in nature. It involves
both resonant and non-resonant components that are inseparable in principle and
always occur together, analgous to those in the complementary physical 
processes of electron-ion scattering and photoionization. 
Experiments or observations of \eion recombination measure the {\it total} 
cross section. Theoretically
therefore a unified treatment with a self-consistent approach is preferable 
to methods that consider
\eion recombination in parts employing different approximations of varying
validity.

Historically, \eion  recombination is
usually considered in two main but separate parts: 
(a) radiative recombination
(RR) i.e.  direct radiative capture and recombination, 
\begin{equation}
 e + X^{n+} \longrightarrow h\nu + X^{(n-1)+}  ,  
\end{equation}
and dielectronic recombination (DR) i.e. indirect capture and recombination 
through autoionizing states,
\begin{equation}
 e + X^{n+} \longrightarrow X^{(n-1)+**} \longrightarrow h\nu +
X^{(n-1)+} ,
\end{equation}
where the intermediate state (indicated by double asterisks) is a
doubly excited state of the (e + ion) system which introduces a resonance.

    The subject of electron-ion recombination has been one of the most
active areas of research in atomic physics for several decades, both
theoretically and experimentally.  Traditionally the two parts, RR and DR, 
are treated independently. The RR cross
sections are obtained in a straightforward manner using detailed
balance (Milne relation) from ground state photoionization cross sections 
computed using relatively simple approximations such as the central
field or the quantum defect method (Burgess \& Seaton, 1960), without
taking account of resonances. On the other hand the theoretical treatment 
of DR has
a long and interesting history (Seaton \& Storey, 1976, Hahn \& Lagattuta
1988).
The main development was the realization by Burgess (1964) that
DR via the infinite series of resonances in the \eion system is an important 
contributor to the total recombination process. The celebrated General Burgess 
Formula was used in many applications, particularly in astrophysical modeling.
More accurate treatments of DR, generally based on the isolated
resonance approximation using the distorted wave method, were later developed 
taking account of physical effects not included in the Burgess  formula,
such as autoionization into excited levels (e.g. Jacobs \etal 1977).
Nussbaumer and Storey (1983) pointed out the importance of low-energy
resonances that might give significant enhancement of the DR rate in the
low-temperature region. Hahn (1985) provided expressions for DR cross
sections for comparison with experiments.
There are many calculations  for
dielectronic recombination (DR) with highly charged ions
(e.g. Pindzola \etal 1990,1992; Badnell \etal 1990), using not only the
distorted wave method also others such as the
saddle point variation method (Mannervik \etal 1997), that yield
good agreement with experimental data for the ions considered.

In recent years a number of pioneering experimental studies have been 
carried out. Experimental measurements of electron-ion recombination cross
sections using ion storage rings exhibit detailed resonance structures
at very high resolution in beam energy (e.g. Wolf \etal 1991, Kilgus \etal
1990,1993, Mannervik \etal 1997, Schippers \etal 1999). The
experiments measure absolute cross sections and therefore provide 
ideal tests
for theoretical methods, as well as the physical effects included in the
calculations.
In light of the new experimental studies however, and given that the
unified method for electron-ion recombination is quite general, it is
desirable to extend the calculations to elicit detailed features for
direct
comparison with the measured cross sections. One of the goals of the
present article
is to demonstrate the accuracy of the method, on par with
the R-matrix treatment of photoionization and electron impact excitation, 
as well as
to study theoretical issues such as relativistic  effects, the
distinction
between close coupling and independent resonance treatments, the
magnitude of the resonant and the non-resonant (background) cross
sections,
relatively sparse near-threshold resonance structures as opposed to the
dense resonances below the Rydberg series limits, radiation damping of
low-lying autoionizing resonances, etc.

   In the present work we describe photoionization,
elctron-ion scattering, and \eion recombination self-consistently
within the framework of the 
close coupling approximation using the R-matrix method. Combining Eqs. (1)
and (2), and invoking detailed balance, we may write

\begin{equation}
 e + X^{n+} \longleftrightarrow X^{(n-1)+**} \longleftrightarrow h\nu +
X^{(n-1)+} ,
\end{equation}

 where photoionization and recombination (resonant $\oplus$ non-resonant) 
proceed inversely in either
direction. The wavefunctions for the (e + ion) system are
obtained with the same eigenfunction expansion for both processes,
enabling a self-consistent treatment in an ab initio manner. 
The R-matrix package of codes has been extended to 
incorporate the theoretical framework described below.

\section{Theoretical Framework}

 The R-matrix method developed by P.G. Burke and associates (Burke and
Robb 1975, Burke and Berrington 1993) provides a
natural and powerful tool for collisional and radiative electron-ion 
processes based on the close coupling or coupled channel (CC) approximation. 
In particular, the R-matrix method 
has been employed for large-scale computations using efficient codes
developed under the Opacity Project (The Opacity Project 1995,1996,
Seaton 1987, Berrington \etal 1987, Seaton \etal 1994),
and the Iron Project (Hummer \etal 1993, Berrington \etal 1995).
Most of these calculations entail photoionization and electron-ion
scattering, the two processes that form the basis of the present
extension of the R-matrix method to unified \eion recombination (Nahar
and Pradhan 1992,1994,1995,1997, Zhang and Pradhan 1997, Zhang \etal 1999).

 In the following subsections we describe the basic theory. 

\subsection{Coupled channel wavefunction and radiative transitions}

The total wavefunction for a (N+1)-electron \eion system in the CC
approximation is described as:
\begin{equation}
\Psi_E(e+ion) = A \sum_i^N \chi_i(ion)\theta_i + \sum_{j} c_j \Phi_j(e+ion),
\end{equation}
\noindent
where $\chi_i$ is the target ion or core wavefunction in a specific state
$S_iL_i\pi_i$ or level $J_i\pi_i$, $\theta_i$ is the wavefunction of the
interacting (N+1)th electron in a channel labeled as
$S_iL_i(J_i)\pi_i \ k_{i}^{2}\ell_i(SL\pi~or~ \ J\pi)$; $k_{i}^{2}$ is the
incident kinetic energy.
$\Phi_j$ is the correlation functions of (e+ion) system that compensates
the orthogonality condition and short range correlation interations.
The complex resonant structures in photoionization, recombination, and in
electron impact excitation are included through channel couplings.

Relativistic effects are included through Breit-Pauli approximation in
intermediate coupling. The (N+1)-electron Hamiltonian in the Breit-Pauli
R-matrix (BPRM) method, as adopted in the Iron Project (Hummer \etal 1993), is
\begin{equation}
H_{N+1}^{\rm BP}=H^{NR}_{N+1}+H_{N+1}^{\rm mass} + H_{N+1}^{\rm Dar}
+ H_{N+1}^{\rm so},
\end{equation}
\noindent
where non-relativisitc Hamiltonian is
\begin{equation}
H^{NR}_{N+1} = \sum_{i=1}\sp{N+1}\left\{-\nabla_i\sp 2 - \frac{2Z}{r_i}
        + \sum_{j>i}\sp{N+1} \frac{2}{r_{ij}}\right\} .  
\end{equation}
\noindent
$H_{N+1}^{\rm mass} = -{\alpha^2\over 4}\sum_i{p_i^4}$ is the mass 
correction term, $H_{N+1}^{\rm Dar} = 
{Z\alpha^2 \over 4}\sum_i{\nabla^2({1\over r_i})}$ is the Darwin term, 
and $H_{N+1}^{\rm so} Z\alpha^2 \sum_i{1\over r_i^3}{\bf l_i.s_i}$ is the 
spin-orbit interaction term.
Spin-orbit interaction splits the LS terms into fine-structure levels
labeled by $J\pi$ where $J$ is the total angular momentum. 
Solutions of the Schrodinger equation, $H^{BP}_{N+1}\Psi = E\Psi$
which becomes a set of coupled equations with the CC expansion, give the
bound wavefunctions, $\Psi_B$, for negative energies (E $<$ 0),
and continuun wavefunction, $\Psi_F$, for positive energies (E $\geq$ 0).

The transition matrix elements for various atomic processes are: 
$<\Psi_B || {\bf D} || \Psi_{F}>$ for photoionization and recombination, 
$<\Psi_B || {\bf D} || \Psi_{B'}>$ for oscillator strength, 
$<\Psi_F | H(e + ion) | \Psi_{F'}> $ for electron impact excitation, 
where {\bf D} is the dipole operator,  ${\bf D}_L = \sum_i{r_i}$ in 
"length" form and ${\bf D}_V = -2\sum_i{\Delta_i}$ in "velocity" form
with the sum over the number of electrons. 
The matrix elements are divided
into inner and outer region components corresponding to the R-matrix
boundary at a suitably chosen radius (r = a), and asymptotic wavefuctions
for r $\longrightarrow \infty$.

The transition matrix element with the dipole operator can be reduced
to the generalized line strength defined, in either length or velocity
form, as
\begin{equation}
S_{\rm L}=
 \left|\left\langle{\Psi}_f
 \vert\sum_{j=1}^{N+1} r_j\vert
 {\Psi}_i\right\rangle\right|^2 \label{eq:SLe},
~~~S_{\rm V}=\omega^{-2}
 \left|\left\langle{\Psi}_f
 \vert\sum_{j=1}^{N+1} \frac{\partial}{\partial r_j}\vert
 {\Psi}_i\right\rangle\right|^2. \label{eq:SVe}
\end{equation}
where $\omega$ is the incident photon energy in Rydberg units, and
$\Psi_i$ and $\Psi_f$ are the wave functions representing the
initial and final states, respectively.

The photoionization cross section ($\sigma_{PI}$) is proportional to
the generalized line strength ($S$),
\begin{equation}
\sigma_{PI} = {4\pi \over 3c}{1\over g_i}\omega S.
\end{equation}
where $g_i$ is the statistical weight factor of the initial state.

\subsection{Electron-ion recombination --- detailed balance}

The total recombination cross section is the sum of cross sections for
recombination into the infinite number of recombined levels
of the \eion system, as illustrated schematically in Fig.~1. 

\begin{figure} %
\psfig{figure=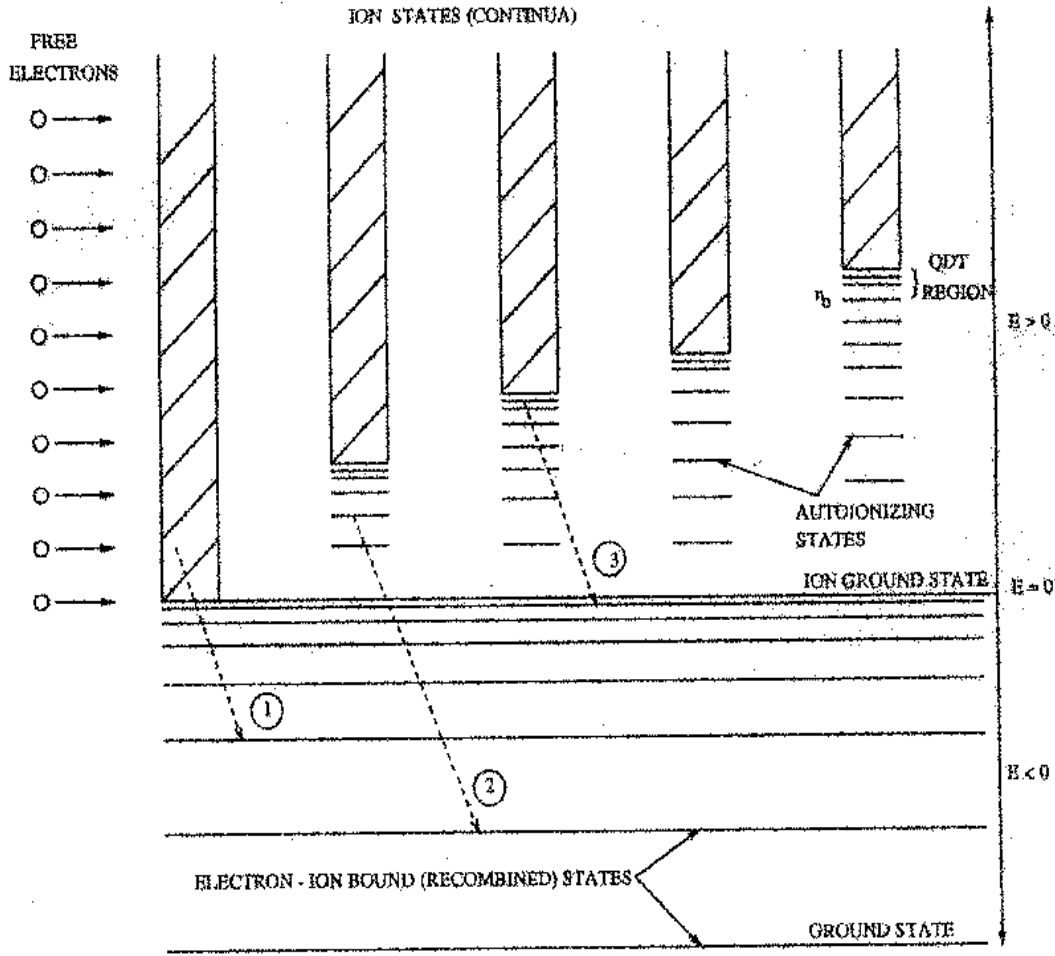,height=14cm,width=15.0cm}
\caption{Schematic energy diagram of unified \eion recombination. The 
infinite series of autoionizing resonances converging onto the various
excited target states are in the positive energy region E $>$ 0, while
recombined states are in the negative energy region E $<$ 0. Broken 
lines with arrows represent photon emission during recombination: (1)
recombination through the ground state continuum of the recombined ion,
(2) through a low-n autoionizing state with possibily large interaction
with the continuua (coupling to excited states gives rise to autoionizing
resonances), and (3) through a high-autoionizing state, with negligible
continuum contribution (DR) only.}
\end{figure}

We divide this infinite set into 3 groups: (i) recombination into low-n 
levels with $n \leq n_o$, (ii) contribution from high-n resonances to 
bound levels with $n_o \leq n \leq \infty$ (DR only), and (iii) 
``top-up" background contribution from $n > n_o$ (RR only). The \no is
chosen such that all three subsets are independent and complementary.
The contributions to recombination from states with $n \leq n_o$ (i)
are obtained from detailed $\sigma_{PI}$, typically with extensive
resonance structures, using the principle of detailed balance (Milne
relation).  The recombination cross section, $\sigma_{RC}$, is related to
photoionization cross section, $\sigma_{PI}$, through the principle of 
detailed balance,
\begin{equation}
\sigma_{RC} = \sigma_{PI}{g_i\over g_j}{h^2\omega^2\over 4\pi^2m^2c^2v^2}.
\end{equation}
The recombination rate coefficient, $\alpha_{RC}$, is obtained as
\begin{equation}
\alpha_{RC}(T) = \int_0^{\infty}{vf(v)\sigma_{RC}dv},
\end{equation}
where $f(v)$ is the Maxwellian velocity distribution function. The total
$\alpha_{RC}$ is obtained from contributions from infinite number of
recombined states as:
\begin{equation}
\alpha_R(T) = \sum_{i_b}{g_i\over g_j}{2\over kT\sqrt{2\pi m^3kc^2T}} 
\int^{\infty}_0E^2\sigma_{PI}(i_b;\epsilon)
e^{-{\epsilon\over kT}}d\epsilon,
\end{equation}
\noindent
where $E = \hbar \omega = \epsilon+I_p$, $\epsilon$ is the photoelectron
energy, and $I_p$ is the ionization potential. The integration over 
$\infty$ energy range of the photoelectron is carried out as described in
(Nahar \& Pradhan 1994). $\sigma_{PI}$, obtained including the autoionizing 
resonances, essentially provide total $\alpha_R(T)$ incorporating both the 
RR and the DR in a unified manner. 

\subsection{Dielectronic recombination via high-\en resonances}

Recombination into the high-$n$ levels of the \eion system via resonances
with $n_o < n \leq \infty$ approaching target ions series limits, 
(process (3) in Fig.~1) is included by 
implementing the extension of precise theory of DR by Bell and Seaton
(hereafter BS 1985, as adapted in Nahar and Pradhan 1994,1995) based on 
the CC approximation.
To each excited threshold $S_iL_i(J_i)\pi_i$ 
of the $N$-electron target ion, there corresponds an infinite series of 
($N$+1)-electron states, $S_iL_i(J_i)\pi_i\nu\ell$, to which 
recombination can occur, where $\nu$ is the effective quantum number of
the autoionizing level. The recombining electron is treated as a
`spectator', recombining from an autoionizing level
$\nu\ell$ to the corresponding \en$\ell$ bound level following
the ion core transition. For sufficiently high-\en 
DR dominates the recombination process and the background 
recombination is negligibly small. The contributions from these
states are added by calculating the collision strengths, $\Omega_{\rm DR}$.
Several aspects related to the application of the theory to the 
calculation of DR collision strengths are described in Nahar and Pradhan
(1992, 1994, 1995), Nahar (1996), Zhang el al. (1999). We sketch below 
a few working expressions derived from the theory.

Including radiative interactions in an {\it ab initio} manner in the
total Hamiltonian for the e+ion system, a generalized electron-photon 
scattering matrix $\cal S$ may be obtained as (Davies \& Seaton 1969)
\begin{equation}
 {\cal S} = \left( \begin{array}{lr}
                   {\cal S}_{\rm ee}  & {\cal S}_{\rm ep} \\
                   {\cal S}_{\rm pe}  & {\cal S}_{\rm pp}
                   \end{array}
            \right)
\end{equation}
where ${\cal S}_{\rm ee}$ is the matrix for electron scattering {\em
including}
radiation damping; ${\cal S}_{\rm pe}$ is the matrix for electron capture
followed by radiative decay with the emission of a photon; ${\cal
S}_{\rm ep}$ that for the inverse process of photoionization; and ${\cal
S}_{\rm pp}$ for photon-photon scattering. In the absence of 
interaction with the radiation field ${\cal S}_{\rm ee}$
is the usual scattering matrix {\bf S}. The unitarity condition for 
(electron + photon) $\cal S$ matrix reflects the conservation of both the 
incident electron and the emitted photon flux (Davies and Seaton 1969), i.e.

\begin{equation}
 {\cal S}_{\rm ee}^{\dagger}{\cal S}_{\rm ee} + {\cal S}_{\rm pe}^{\dagger}{\cal
S}_{\rm pe} = 1.
\end{equation}

The electron-electron scattering matrix, ${\cal S}_{\rm ee}$, may
be again partitioned into sub-matrices of open and closed channels, in
the energy region  below threshold, in terms of its analytic 
continuation given by 
the matrix {\bf $\chi$} as ${\bf \chi_{oo}, \chi_{oc},
\chi_{co}},$ and ${\bf \chi_{cc}}$, where `o' denotes the open and
`c' the closed channels. The open channels are those that are accessible
to the incident electron for excitation of a target state in that
channel; a closed channel refers to electron energies below an
inaccessible target threshold. A given Rydberg series of resonances,
\unboldmath
converging on to a target threshold $S_tL_t$, corresponds  to the closed
channel $(S_tL_t) \epsilon\ell$, where  $\epsilon = -1/\nu^2$, $\nu$
is the effective quantum number associated with the resonance series. The 
scattering matrix, {\boldmath ${\cal S}_{ee}$}, is then obtained as 
(BS 1985)
\begin{equation}
\mbox{\boldmath ${\cal S}_{ee} =  \chi_{oo} - \chi_{oc}[\chi_{cc}$} -
g(\nu){\rm exp}(-2{\rm i}\pi\nu)]^{-1} \mbox{\boldmath $\chi_{co}$},
\end{equation}
where $ g(\nu) = {\rm exp}(\pi\nu^3\gamr /z^2) $; $\gamr$ is the sum of 
all possible radiative decay probabilities for the resonance series. These 
decay probabilities correspond to radiative transitions within the ion 
core. The outer electron is treated as a ``spectator'', in a high-$n$ 
resonance state, interacting but weakly with the core.

The electron flux trapped in the closed channel resonances may decay
radiatively to bound states of the e+ion system. In multi-channel quantum
defect theory we diagonalize the {\boldmath $ \chi$} matrix as 
$\mbox{\boldmath $\chi_{cc}N = N$}\chi_{cc}$,
where $\chi_{cc}$ is a diagonal matrix and {\boldmath $N$} is the diagonalizing
matrix with {\boldmath $ N^{\rm T}N = 1$}. In terms of {\boldmath $N$} we write 
{\boldmath $\chi_{oc}^{\prime} = \chi_{oc}N$} and
{\boldmath $\chi_{co}^{\prime} = N^{\rm T}\chi_{co}$}, 
where {\boldmath $N^{\rm T}$} is the
transpose of {\boldmath $N$}. The DR probability, 
for an entrance or incident open channel $\alpha$, is obtained from the 
unitarity condition as
\begin{equation}
 P_{\alpha}({\rm DR}) = \mbox{\boldmath $(1 - {\cal S}_{ee}^{\dagger}{\cal
S}_{ee})_{\alpha\alpha}$}.
\end{equation}
Substituting the proper expressions, the DR probability can be written
as (Nahar and Pradhan 1994, Nahar 1996)
\begin{eqnarray}
P_\alpha = G(\nu)\sum_\gamma\left\{
\left(\sum_{\gamma'}\chi'_{\alpha\gamma'}{\bf N}_{\gamma 
\gamma'}\right)\left[{1\over {\bf \chi}_{\gamma \gamma}-g(\nu)
{\rm exp}(-2\pi {\rm i}\nu)}\right] \right. \cr
\times \left[{1\over {\bf \chi}_{\gamma 
\gamma}^{\ast}-g(\nu){\rm exp}(+2\pi {\rm i}\nu)}\right]
\left. \left(\sum_{\gamma'}{\chi'}_{\gamma' \alpha}^\ast {\bf N}_{\gamma
 \gamma'}^\ast\right)\right\},
\end{eqnarray}
\noindent
where $G(\nu) = g(\nu)^2 - 1 = {\rm exp}(2\pi\nu^3\gamr /z^2)$ - 1.
The summations go over the closed channels $\gamma\gamma'$
contributing to DR. The sum over the diagonal elements of all open channels
linked to the ground state of the target ion gives the probability of DR 
through radiative transitions between the excited states and the ground state.
As we are interested only in the detailed DR collision strengths, expressions 
derived for resonance averaged DR collision strengths, that are useful 
in the calculation of recombination rate coefficients, are not given here 
but may be found in Nahar and Pradhan (1994), Nahar (1996).

The DR collision strength, $\Omega(DR)$, is obtained as
\begin{equation}
\Omega(DR) = \sum_{SL\pi} \sum_\alpha{1\over 2}(2S+1)(2L+1)P_\alpha^{SL\pi}.
\end{equation}
The $\Omega_{\rm DR}$ are calculated, in a self-consistent manner, 
using the same CC wavefunction expansion that is 
used for the calculation of $\sigma_{\rm PI}$. The DR cross section, in
Megabarns (Mb), is related 
to the collision strength, $\Omega_{\rm DR}$, as
\begin{equation}
\sigma_{\rm DR}(i\rightarrow j)({\rm Mb}) = \pi \Omega_{\rm DR}(i,j)/(g_ik_i^2)
(a_o^2 \times 10^{18}) ,
\end{equation}
where $k_i^2$ is the incident electron energy in Ry.

\subsection{Radiation damping of low-\en resonances}

The dominant DR contribution from the energy region below series limits due to 
radiation damping of high-\en resonances of group (ii), 
with $n_o < n \leq \infty$,  is
considered using the BS theory as described above. Generally, for
low-\en resonances of group (i) we do not expect radiation damping to be
an important consideration since the autionization rates are much larger
than radiative decay rates. However, in some cases the effect could be
significant and the low-\en resonance structures in the detailed
photoionization cross sections need to be resolved and damped.

Although radiation damping of resonances in electron-ion interactions
has been studied for a long time,
the effect has been found to be
important only for highly charged H-like and He-like ions where
the 2p $\rightarrow$ 1s and the
1s2p $^1{\rm P}^o \rightarrow {\rm 1s}^2 \ ^1$S
dipole transitions respectively,
correspond to radiative decay probabilities $\Gamr$,
of the order of $10^{13}-10^{14} \ {\rm s}^{-1}$, approaching
typical autoionization rates $\Gama$.
Employing the branching ratio between autoionization and
radiative decay channels (Bates and Massey 1943,
Burgess 1964, Presnyakov and Urnov 1975), Pradhan (1981) showed
DR may result in a significant reduction in
the autoionization contribution to
electron-ion scattering cross section. It was further pointed out
that the flux lost from electron impact excitation (EIE) is equal to the
DR contribution; detailed calculations were presented for He-like
O{\sc\,VII} and Fe{\sc\,XXV}.
Calculations for rate coefficients however showed that the net effect
was not large. In a combined study of fine structure, autoionization,
and radiative decays of {\it high-n} resonances up
to the Rydberg series limits at target thresholds,
Pradhan (1983a,b) found maximum reductions in the
EIE rate coefficients for the metastable
$1\ ^1{\rm S}_0 - 2\ ^3{\rm S}_1$ transition
of 9\% for He-like Fe{\sc\,XXV} and 19\% for Mo{\sc\,XLI}.
The effect on EIE of the radiative decay of {\it low-n} resonances is
small. Following Pradhan's work, Trefftz (1983) studied several ionic systems
and found that ``this
effect is important only in certain cases of which highly charged
He-like ions are the most prominent".

 Radiative decay of low-\en resonances in photoionization cross
sections, relevant to \eion recombination, is studied using two methods. 
The first,
due to Sakimoto \etal (1990), entails fitting the dipole matrix element
to a form
\begin{equation}
D(E)=D^0(E) + {A \over E-Z^*};\ \ Z = E_0 - {{\rm i} \over 2} \Gama
\end{equation}
where $E_0$ is the resonance position and $\Gama$
is the autoionization width in Ry. The radiative decay width $\Gamr$
is then obtained by 
\begin{equation}
\Gamr = 4\pi^2{\mid A \mid^2 / \Gama}.
\end{equation}
The {\it second order} radiative effects can then be included by considering
\begin{equation}
 D(E) \longrightarrow {D(E) \over 1+L(E)}, 
\end{equation}
where, according to the BS theory, the operator L(E) is given by
\begin{equation}
L(E) = \pi^2\mid D^0(E)\mid^2 + 2\pi^2{A^*D^0(E) \over E-Z}
+2\pi^2{\mid A \mid^2 \over (E-Z)(Z-Z^*)}.
\end{equation}
A version of the BP R-matrix (BPRM) codes has been employed 
(Eissner W, to be submitted to {\it Comput. Phys. Commun.}); 
$\Gamr$ are obtained in
intermediate coupling from the BPRM calculations.

Resonances and radiation damping in
photoionization and photo-recombination may also be considered in the
relativistic distorted-wave (RDW) approximation (Sampson and Zhang
1995) accurate for highly charged ions. Using
oscillator strengths or the radiation
rates calculated by the Dirac-Fock-Slater (DFS) structure
program (Sampson \etal 1989), and autoionization
rates by the RDW program,
the photo-excitation cross section from a level $k$ to
a doubly-excited level $j$
is given by the absorption radiative rate $A^r_{kj}=\hbar \Gamr_{kj}$
\begin{equation}
\sigma_{kj} = {4\pi^2a_0^2 \over \alpha^2E^2} \Gamr_{kj}
\delta(E-E_{jk}),
\end{equation}
where $E_{jk}$ is the transition energy, $a_0$ is the Bohr radius and
$\alpha$
is the fine-structure constant. Replacing the $\delta$ function by a
Lorentz
profile, and from detailed balance with $g_j\Gamr \equiv g_j
\Gamr_{jk} = g_k \Gamr_{kj}$, we obtain the photo-recombination
cross section for the transition from level
$i$ to level $k$ without radiation damping
\begin{equation}
\sigma^{\rm PR}_{ik} = \sum_j {\pi^2a_0^2 \over E} {g_j \over g_i} \Gamr
\ {\Gama/2\pi \over (E-E_{jk})^2+(\Gama /2)^2},
\end{equation}
where $g$ is the statistical weight. When radiation damping is
included the autoionization width $\Gama$
should be replaced in the denominator by the total width
$\Gama+\Gamr$ in the Lorentz profile, and we have
\begin{equation}
\sigma^{\rm PR}_{ik} = \sum_j {\pi^2a_0^2 \over E} {g_j \over g_i} \Gama
\ {\Gamr \over \Gama+\Gamr}
\ {(\Gama+\Gamr)/2\pi \over (E-E_{jk})^2+(\Gama/2+\Gamr / 2)^2}.
\end{equation}
This is the usual equation for the di-electronic recombination cross
section
such as that used in the RDW calculations (Sampson and Zhang 1995),
$\Gamr /(\Gama+\Gamr)$ being the branching ratio.

In a detailed study Pradhan and Zhang (1997) showed that the fitting procedure
is accurate and also provides a check for
the version of the BPRM codes used.
They found approximately 10-20\% agreement between calculated 
dielectronic-resonance 
strengths for the KLL group of
resonances in recombination with Fe~XXV, and the corresponding 
values derived
experimentally by Beiersdorfer \etal (1992), and other calculations.

In general our studies have shown that with the exception of ions with
H-like and He-like core transitions,  radiative decay of low-\en
resonances do not show an appreciable effect on effective recombination
cross sections (except in isolated or small energy ranges) 
or rate coefficients (e.g. Zhang \etal 2001).

\subsection{High-\en background recombination --- ``top-up"}

At very low electron energies and temperatures the recombination 
is dominated by radiative recombination to very high-n states just below 
the ground state of the recombining ion. These contributions are derived 
as discussed in Nahar (1996). They are included in the hydrogenic 
approximation. Although the 
recombination to high-n states is dominated by DR at higher temperatures,
the radiative part involving the photoionization cross section of
high-n states is "topped-up" with the hydrogenic recombination rate
coefficient. For an ion with charge z, we have the z-scaled formula 
$\alpha_R(z,T)$ = $\alpha_R(1,T/z^2)$, in terms of the recombination
rate coefficient for neutral hydrogen. We calculate the
hydrogenic recombination rate coefficients, $\alpha_R(z,T)$, for states
n = 10 to 800 employing photoionization cross sections of hydrogen obtained
using the FORTRAN program by Storey and Hummer (1992), and for states
n = 801 to $\infty$ using the sum (Hummer 1994)
\begin{equation}
\Delta(n) = \alpha_n\left(n\over n+1\right)^3 \left(1+n\over 2\right)
\end{equation}

\section{Results and discussion}

 We present a wide range of results for ions of varying complexity to
demonstrate the generality of the self-consistent unified method, and
physical interrelationships among complementary processes of
photoionization, recombination, and excitation.

\subsection{Comparison with experiments}

 As a test of the accuracy and resolution of the \eion recombination
cross sections and photoionization cross sections using the
self-consistent approach based on the R-matrix method, we have carried
out a number of detailed comparisons for both recombination and
photoionization with available experiments discussed in the following
section.

\subsubsection{Unified \eion recombination cross sections}

 Very high resolution experimental cross sections are now being measured
for \eion recombination on heavy ion synchrotron storage rings, the Test
Storage Ring (TSR) in Heidelberg, and CRYRING in Stockholm.
 Experimental verification of the unified \eion recombination
cross sections, with those measured in detail on ion storage rings,
has been done for several
(recombined) ions for which experimental data is available: 
C~III (Mannervik \etal 1998, Schippers \etal 2001),
C~IV,C~V,O~VII (Zhang \etal 1999), Ar~XIII (Zhang and Pradhan 1997),
Fe~XVII (Savin \etal 1999). 

 Fig. 2 demonstrates recombination to Ne-like Fe~XVII (Pradhan \etal
2001, Zhang \etal 2001) with
a BPRM calculation that includes only the first three levels $2s^22p^5
(^2P^o_{3/2,1/2}),2s2p^6(^2S_{1/2}$ levels in the eigenfunction expansion
of the recombining ion F-like Fe~XVIII. The main DR contributions arise
from the two $\Delta n$ = 0 dipole transitions within the core ion.
The computed rate coefficients 
agree with the sum of RR and experimentally derived DR rate coefficients
(Savin \etal 1999) to within 20\%. A more extensive
multi-configuration expansion up to the $n$ = 3 levels of
Fe~XVIII, with 5 spectroscopic configurations
$2s^2p^5, 2s2p^6, 2s^22p^4 \ (3s, 3p, 3d)$, is in progress. These
calculations will also include the $\Delta n = 1 $ dipole transitions that
give rise to large resonances structures in the \en = 2 to 3
range, as found by Zhang \etal (2001). 

\begin{figure} %
\psfig{figure=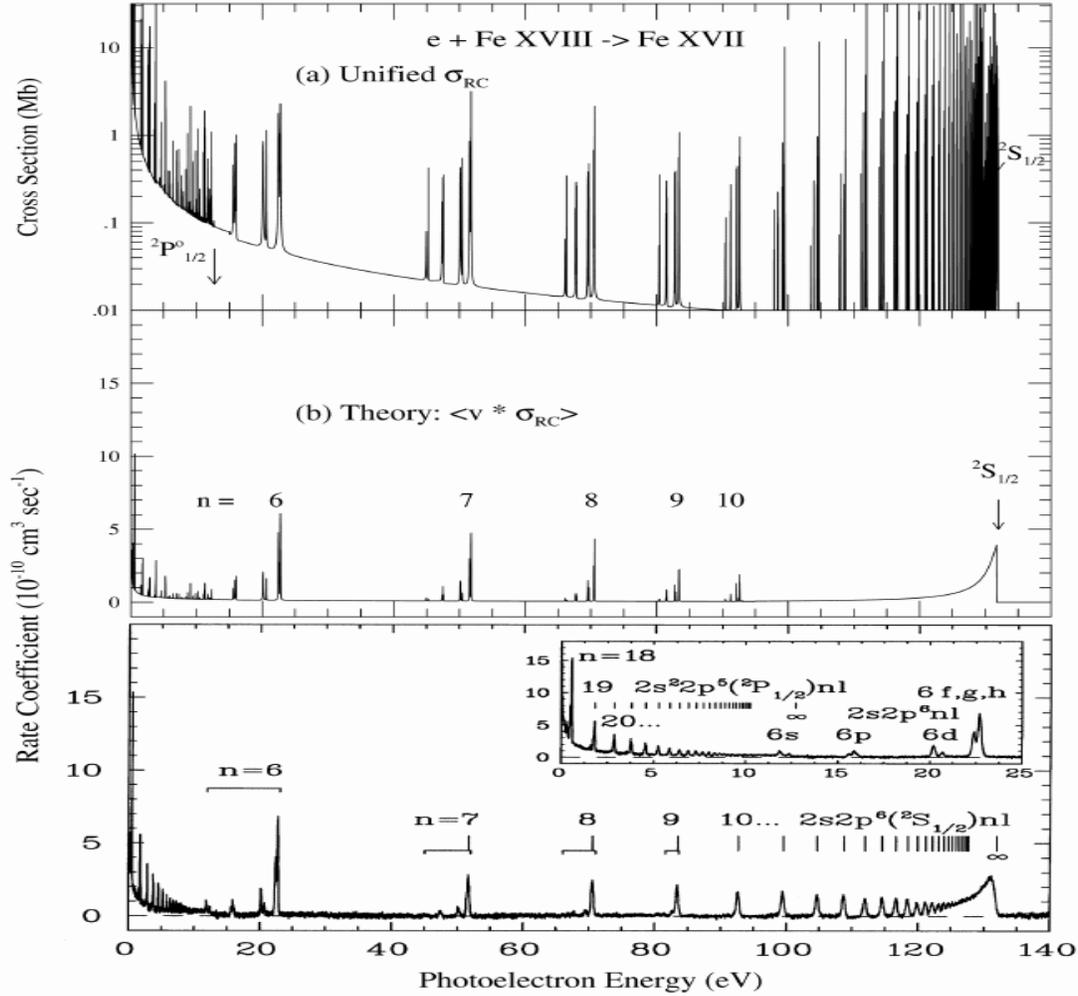,height=14cm,width=15.0cm}
\caption{Unified (e + Fe~XVIII) $\rightarrow$ Fe~XVII recombination
cross sections (upper panel) with detailed resonance complexes below the
n = 2 thresholds of Fe~XVIII (Pradhan \etal 2001); gaussian
averaged over a 20 meV FWHM (middle panel); experimental data from ion
storage ring measurements (bottom panel, Savin \etal 1999).}
\end{figure}

 Another set of relativistic close coupling BPRM calculations (Pradhan
\etal 2001) compared the unified \eion recombination cross sections for
recombination from Li-like C~IV to Be-like C~III with two different 
ion storage experiments, on
CRYRING (Mannervik \etal 2001) and TSR (Scippers \etal 2001). Of particular
interest in this case was the presence of large autoionizing $2p4\ell$ 
resonances in the near-threshold region. The effective integrated
value of the unified cross sections over this resonance complex 
lies between the two sets of experimental values, and agrees with both
to about 15\%. Fig. 3 shows the detailed comparison of the BPRM results
with the TSR data over the region covered by the experiement. 
Schippers \etal had earlier found that the previous LS coupling 
calculations for carbon and nitrogen ions (Nahar and Pradhan 1997) gave
total C~III rate coefficients that agreed with the sum of RR and
experimental DR rate coefficients to within experimental uncertainties
for all temperatures T $>$ 5000 K. The discrepancy at lower temperatures
was due to the $2p4\ell$ complex, as discussed in detail by Mannervik
\etal (1998) and Pradhan \etal (2001), and resolved by the higher accuracy 
BPRM calculations shown in Fig. 3.

\begin{figure} %
\psfig{figure=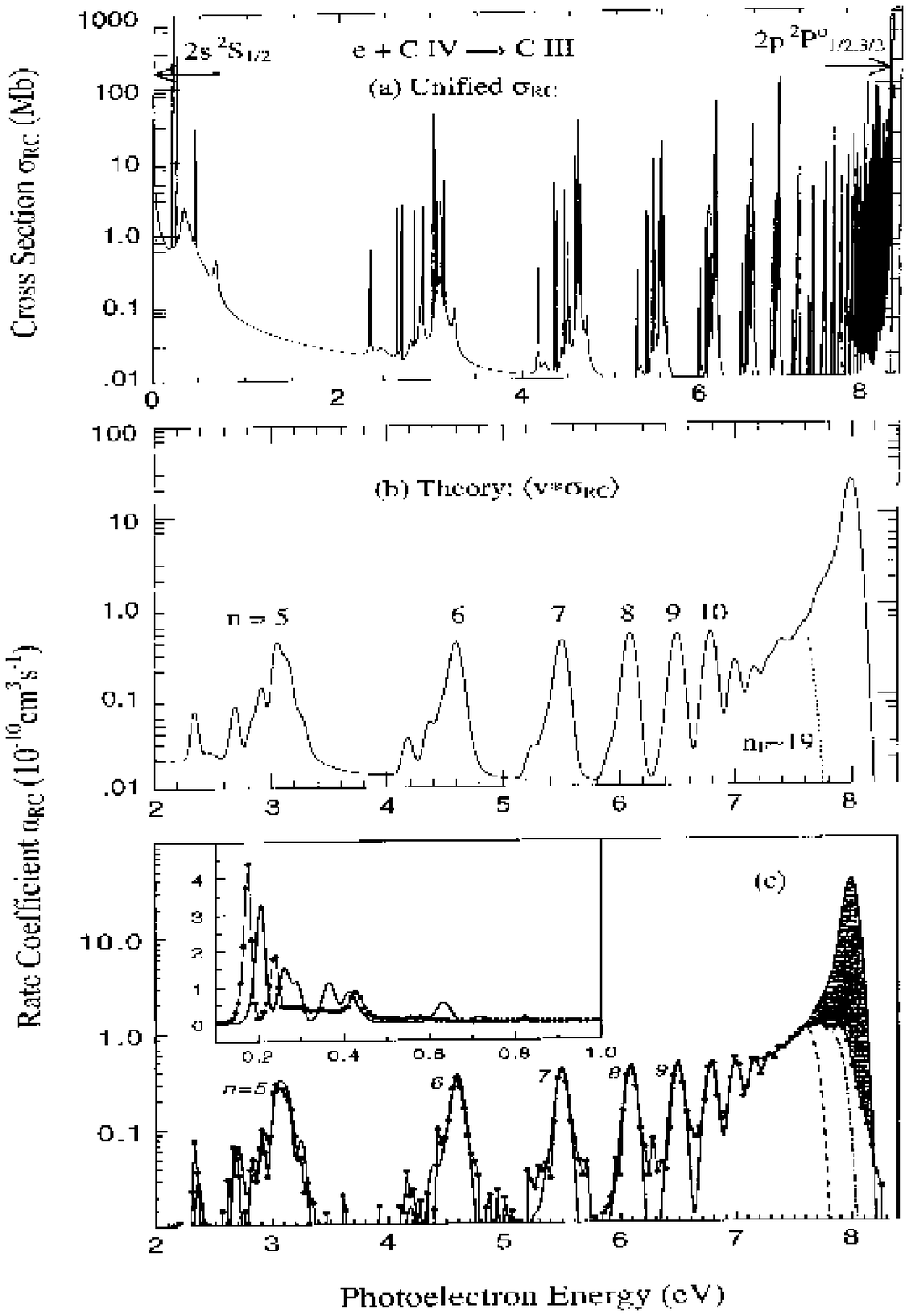,height=14cm,width=15.0cm}
\caption{(a) Unified (e~+~C~IV) $\longrightarrow$ C~III 
recombination cross section $\sigma_{RC}$
with detailed resonance structures (Pradhan \etal 2001); (b) theoretical 
rate coefficient (v $ \cdot \sigma_{RC}$) convolved
over a gaussian with experimental FWHM at the Test Storage Ring (TSR)
(Scippers \etal (2001); (c) the experimentally
measured rate coefficient. The unified $\sigma_{RC}$ in
(a),(b) incorporate the background cross section eliminated from the
experimental data in (c). The dashed and dot-dashed lines represent
approximate field ionization cut-offs.}

\end{figure}

\subsubsection{Photoionization cross sections}

New advances have been made recently in measurements of photoionization cross
sections with unprecedented resolution by three experimental groups:
the University of Nevada, Reno, with the Advanced Light Source
at Berkeley, Aarhus University, and University of Paris-Sud.
All three experimental groups have compared their measured data with
theoretical R-matrix calculations for several carbon and oxygen ions 
(e.g. Kjeldsen \etal 1999, Covington \etal 2001, Nahar 2003), as
well as for heavier systems such as Fe~II (Kjeldsen \etal 2002) 
and Fe~IV (R. Phaneuf \etal, 
in progress). Fig. 4 shows a sample comparison between theory and
experiment.

 Of great interest is the comparison between theory and experiments
to ascertain the mixture of ions in the ground
and metastable levels in the experimental beam. The signature of
metastable levels manifests itself in experimental data as resonances
that appear {\it below} the ground state ionization threshold, since the
metastable levels are photoionized at lower energies. Theoretical
photoionization cross sections for the ground and metastable levels are
therefore weighted and combined in order to compare with and interpret
experimental measurements. This is likely to be of considerable
importance in practical applications in laboratory and astrophysical
plasmas where metastable levels may be significantly populated.

\begin{figure} %
\psfig{figure=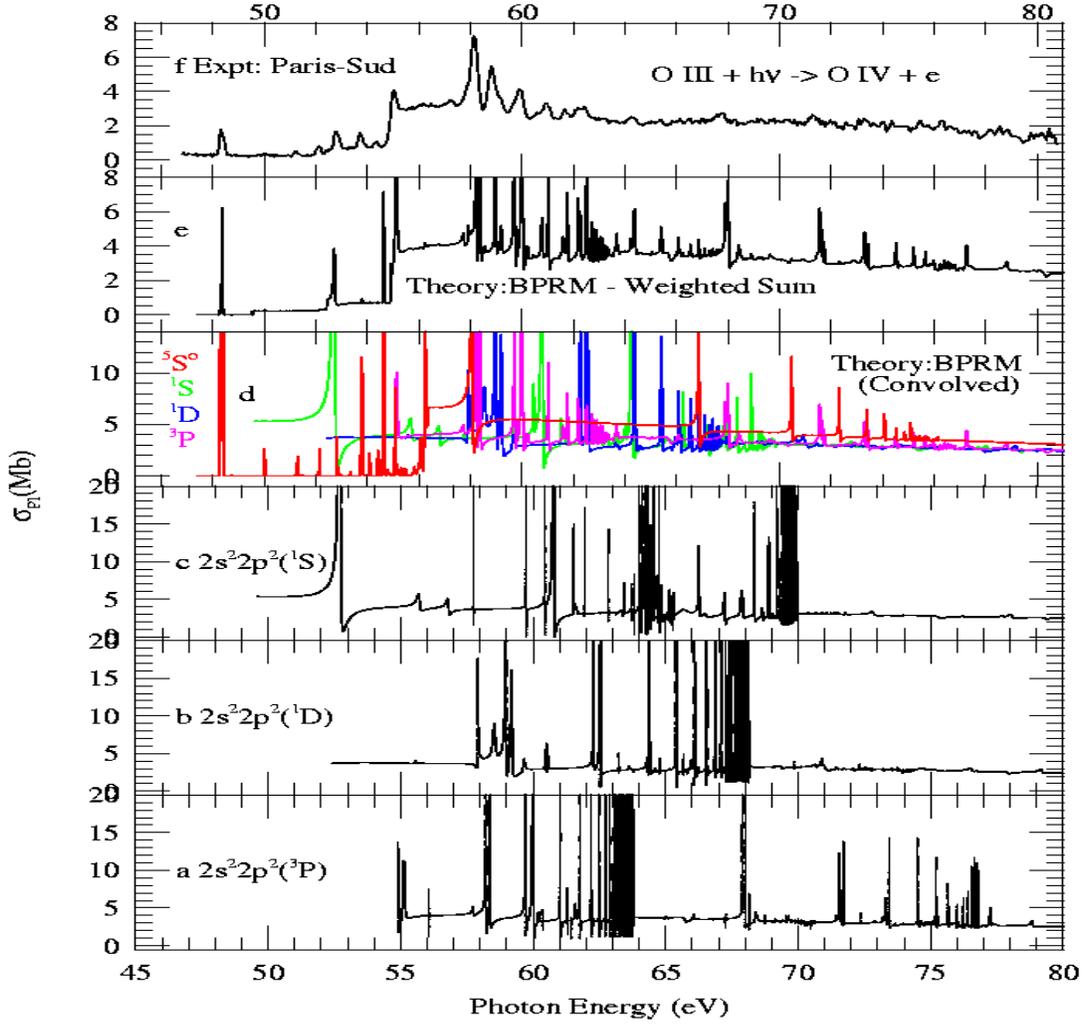,height=14cm,width=15.0cm}
\caption{Photoionization cross sections $\sigma_{PI}$ of O III (Nahar
2003): Panels a, b, c - $2s^22p^2(^3P,^1D,^1S)$ states respectively; 
d - convolved cross sections of $^3P,^1D,^1S,^5S^o$ states over the
experimental beam distribution; e - weighted sum of convolved
cross sections; f - experimental cross sections measured 
at the University of Paris-Sud by Champeaux et al. (2003).}
\end{figure}

 The level of sophistication of the state-of-the-art theoretical and
experimental works on photoionization and \eion recombination 
is evident from the examples in this section (and cited references).
However, it needs to be emphasized that while theoretical calculations
are capable of matching experimental accuracy and resolution, and vice versa, 
some physical effects and processes deserve careful consideration, 
as discussed later.

\subsection{Strongly coupled systems: Iron ions}

 The CC approximation is especially designed for strongly coupled
systems. Among the most difficult R-matrix calculations are those for low
ionization stages of iron. Hitherto, most of the work has been done in
LS coupling since BPRM calculations are, as yet, intractable unless
their scope is quite limited. In several previous calculations, 
the unified method has been applied to photoionization and recombination 
cross sections and rate coefficients for Fe~I --- V (see Bautista and
Pradhan for references). The CC expansions include 
all LS terms of the ground and excited even parity configuration of the 
recombining ion, and
the first excited odd parity configuration that enables dipole
transitions in the core.

The photoionization cross sections, $\sigma_{PI}$, are calculated
including
autoionizing resonances that can enhance the background cross sections
considerably. Fig.~5 shows the photoionization cross sections of the
gound states of Fe~I to Fe~V. Extensive
resonances dominate the cross sections for these complex ions. The
differences with previous calculations, indicating resonance enhancement
due to extensive channel couplings, are up to three orders of magnitude for Fe
I, over an order or magnitude for Fe II, and $\sim$ 50\% for Fe III.
The primary reason for the differences with simpler approximations, such
as the central field approximation, that neglect channel couplings, is
due to the dominant role of the $3d$-shell in photoionization of these
Fe ions, relative to the much smaller cross section of the outer $n = 4$
electrons with increasing energy. CC calculations are therefore
essential to obtain accurate cross sections for these Fe-ions, and indeed for
all neutral and low ionization states of elements. It is remarkable
however that the earlier many-body perturbation theory
calculations for Fe~I by Kelly (1972, dashed line in Fig. 5) show some
similar structures as the R-matrix cross sections.

\begin{figure} %
\psfig{figure=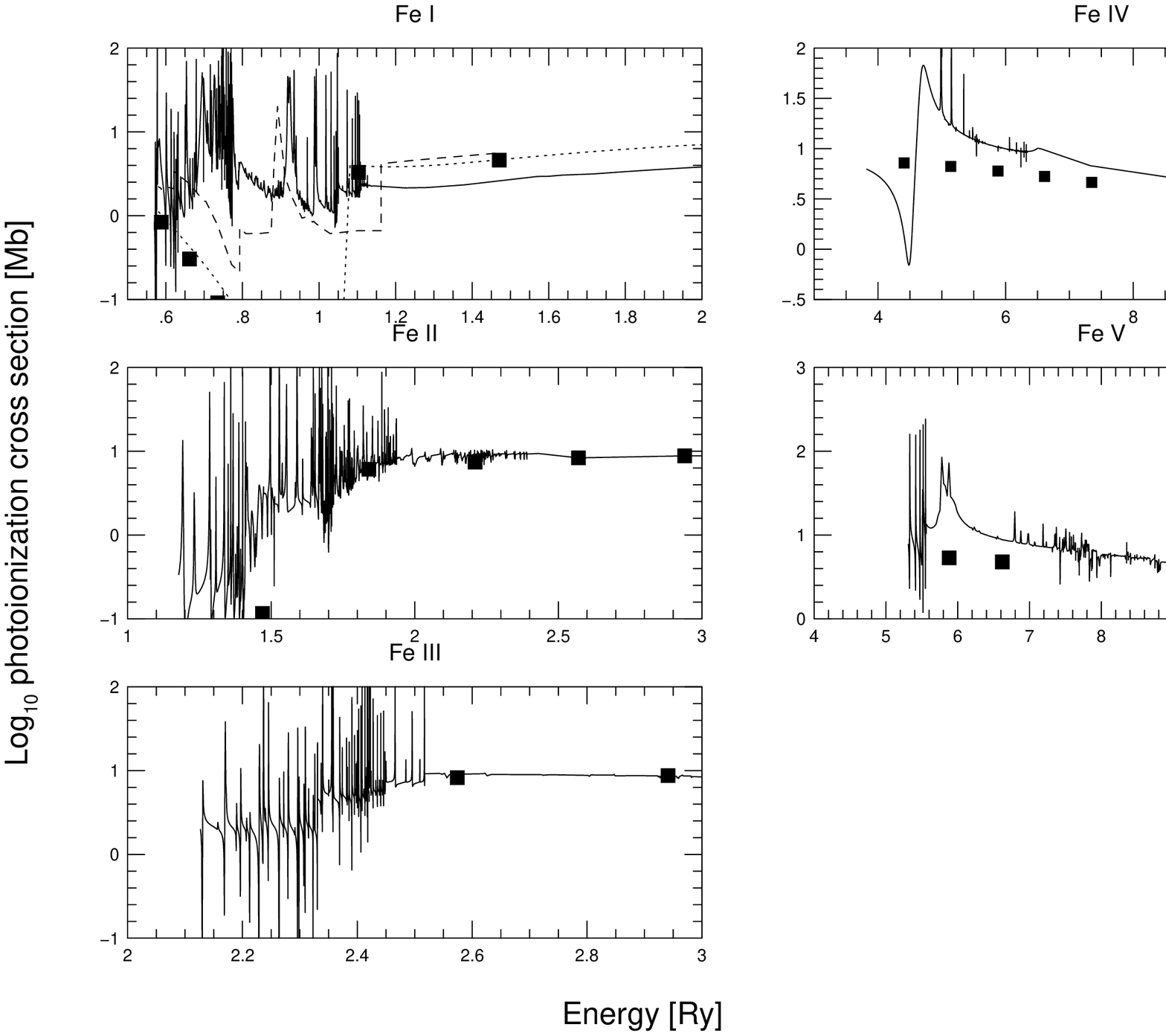,height=9cm,width=15.0cm}
\caption{Photoionization cross sections, $\sigma_{PI}$, of the ground
states of Fe I - Fe V showing detailed autoionization resonances. Also
plotted are the results of Reilman and Manson (1979, filled squares)m
Verner \etal (1993, dotted line), and Kelly (1972, dashed line in Fe~I).}
\end{figure}

 The recombination rate coefficients for Fe~I~-~V are shown in Fig. 6.
These were computed using
photoionization cross sections for the ground state (Fig. 5), and
typically a few hundred excited bound states. The CC expansion for the
recombining ion included target states of configurations denoted as
$3d^q (4s,4p,4d)$. For instance the target expansion for photoionization
and recombination of Fe II: (e~+~Fe~III) $\longleftrightarrow$ Fe~II +
h$\nu$, employ an 83-LS term expansion for core ion Fe~III with
configurations $3d^6, 3d^5 (4s,4p,4d)$ (Nahar and Pradhan 1994). 
The low-energy resonance structures are well represented, 
as seen in Fig. 5, including the dipole $3d \longrightarrow 4p$ transitions 
responsible for DR. Therefore at low-temperatures $T < 10^5$ K the
unified \art should be quite accurate. We find considerable differences
with previous works on RR and DR of Fe ions. For example, the Fe~I
the unified recombination rate is up to a factor of 4 higher at low
temperatures than the sum of RR and DR rate coefficients from Woods
\etal (1981).

 Owing to the size of the computational
problem, the much higher energy 
$3d \longrightarrow 4f$ core transitions were not included in the
CC expansion for the target ions. This likely accounts for the
discrepancy in the higher temperature range from the DR bump in \art.
In fact there are two competing effects: enhancement due to the $3d
\longrightarrow 4f$ DR, and reduction due to autoionization into 
many excited states present below the high-lying $3d^q4f$ levels. 
More extensive calculations are needed to fully study this issue.
Fine structure also needs to be considered in order to obtain higher
accuracy low-energy cross sections and \art.

\begin{figure} %
\psfig{figure=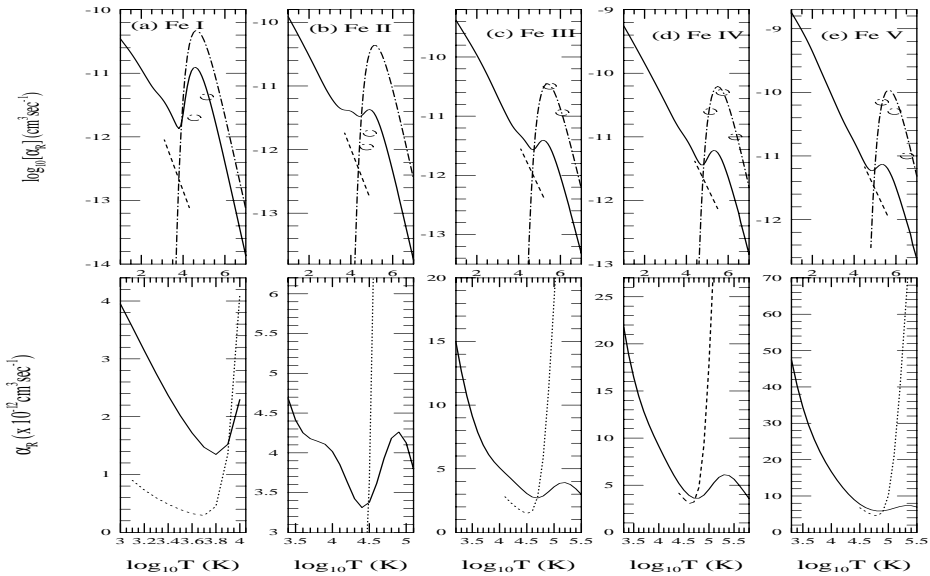,height=9cm,width=15.0cm}
\caption{Comparison of the unified total recombination rate coefficients
($\alpha_R(T)$ (solid lines) for Fe~I (Nahar \etal 1997), Fe~II (Nahar 1997), 
Fe~III (Nahar 1996), Fe~IV (Nahar et al 1998), and Fe~V (Nahar and Bautista
1999). The unified rate coefficeents are compared with the sum of RR and
DR rate coefficients from Woods \etal (1981, dashed lines).}
\end{figure}

 Table 1 gives the unified \art for Fe~I-~V in the low-temperature region
where they are usually abundant in plasmas. Also given in Table 1 are
sample \art from recent works for some other iron ions. 
The unified method, based on the CC approximation, is possibly the only method
capable of accurate results for neutral and low ionization states of elements,
where a separate treatment of RR and DR is unphysical and inaccurate. 
The R-matrix method for \eion recombination is of course equally valid and
accurate for highly charged ions where a separation between the
background and resonances is easier, and the sum of individual RR and DR 
rate coefficients may be of sufficient accuracy. Table 1 also gives the
total unified \art for a few highly ionized iron ions, displaying the
full range of ionization states from Fe~I to Fe~XXV (rate coefficients
for H-like Fe~XXVI are given for completeness and comparison).

\begin{table*}
\caption{Total recombination rate coefficients, $\alpha_R(T)$ (in units
of $cm^3s^{-1}$) for Iron ions in the temperature range of $1.0 \leq 
log_{10}(T) \leq 9.0$.
\label{table3}}
\begin{center}
\scriptsize
\begin{tabular}{clllllllll}
\hline
$log_{10}T$ & \multicolumn{7}{c}{$\alpha_R(T)$} \\
& Fe I & Fe II & Fe III & Fe IV & Fe V & Fe XIII & Fe XXIV & Fe XXV & Fe XXVI \\
\hline
 1.0& 3.48-11&1.23-10&4.20-10&5.57-10&1.80-9 &1.84-8 &1.76-8 &2.80-8 &3.21-8 \\
 1.1& 3.14-11&1.07-10&3.67-10&4.88-10&1.58-9 &1.63-8 &1.57-8 &2.49-8 &2.85-8 \\
 1.2& 2.84-11&9.34-11&3.22-10&4.28-10&1.39-9 &1.44-8 &1.39-8 &2.21-8 &2.54-8 \\
 1.3& 2.58-11&8.13-11&2.82-10&3.74-10&1.22-9 &1.27-8 &1.23-8 &1.97-8 &2.25-8 \\
 1.4& 2.33-11&7.06-11&2.47-10&3.27-10&1.07-9 &1.12-8 &1.09-8 &1.75-8 &2.00-8 \\
 1.5& 2.09-11&6.13-11&2.16-10&2.86-10&9.34-10&9.81-9 &9.68-9 &1.55-8 &1.78-8 \\
 1.6& 1.87-11&5.32-11&1.89-10&2.49-10&8.12-10&8.59-9 &8.56-9 &1.38-8 &1.58-8 \\
 1.7& 1.67-11&4.61-11&1.65-10&2.17-10&7.03-10&7.50-9 &7.57-9 &1.22-8 &1.40-8 \\
 1.8& 1.48-11&3.98-11&1.44-10&1.89-10&6.05-10&6.53-9 &6.68-9 &1.08-8 &1.24-8 \\
 1.9& 1.31-11&3.45-11&1.25-10&1.65-10&5.18-10&5.67-9 &5.88-9 &9.55-9 &1.10-8 \\
 2.0& 1.16-11&2.97-11&1.08-10&1.43-10&4.40-10&4.89-9 &5.17-9 &8.44-9 &9.71-9 \\
 2.1& 1.02-11&2.56-11&9.30-11&1.24-10&3.72-10&4.21-9 &4.55-9 &7.46-9 &8.58-9 \\
 2.2& 9.02-12&2.21-11&7.99-11&1.08-10&3.13-10&3.60-9 &3.99-9 &6.58-9 &7.58-9 \\
 2.3& 7.97-12&1.90-11&6.84-11&9.38-11&2.63-10&3.07-9 &3.49-9 &5.80-9 &6.69-9 \\
 2.4& 7.04-12&1.63-11&5.83-11&8.13-11&2.21-10&2.63-9 &3.06-9 &5.11-9 &5.90-9 \\
 2.5& 6.27-12&1.40-11&4.95-11&7.06-11&1.86-10&2.26-9 &2.67-9 &4.50-9 &5.20-9 \\
 2.6& 5.65-12&1.21-11&4.19-11&6.11-11&1.57-10&1.97-9 &2.33-9 &3.95-9 &4.57-9 \\
 2.7& 5.15-12&1.04-11&3.54-11&5.29-11&1.32-10&1.77-9 &2.03-9 &3.47-9 &4.02-9 \\
 2.8& 4.73-12&9.03-12&2.98-11&4.57-11&1.12-10&1.61-9 &1.76-9 &3.05-9 &3.53-9 \\
 2.9& 4.34-12&7.91-12&2.50-11&3.95-11&9.40-11&1.49-9 &1.53-9 &2.67-9 &3.10-9 \\
 3.0& 3.95-12&6.97-12&2.10-11&3.41-11&7.92-11&1.38-9 &1.33-9 &2.34-9 &2.72-9 \\
 3.1& 3.55-12&6.20-12&1.77-11&2.93-11&6.67-11&1.28-9 &1.15-9 &2.05-9 &2.39-9 \\
 3.2& 3.14-12&5.57-12&1.49-11&2.53-11&5.62-11&1.19-9 &9.93-10&1.79-9 &2.09-9 \\
 3.3& 2.75-12&5.07-12&1.26-11&2.18-11&4.75-11&1.12-9 &8.57-10&1.57-9 &1.83-9 \\
 3.4& 2.37-12&4.68-12&1.07-11&1.88-11&4.03-11&1.08-9 &7.39-10&1.37-9 &1.60-9 \\
 3.5& 2.02-12&4.41-12&9.14-12&1.64-11&3.44-11&1.06-9 &6.35-10&1.20-9 &1.40-9 \\
 3.6& 1.72-12&4.25-12&7.86-12&1.45-11&2.95-11&1.06-9 &5.46-10&1.04-9 &1.23-9 \\
 3.7& 1.47-12&4.18-12&6.86-12&1.28-11&2.56-11&1.06-9 &4.68-10&9.11-10&1.07-9 \\
 3.8& 1.35-12&4.14-12&6.11-12&1.15-11&2.22-11&1.05-9 &4.01-10&7.94-10&9.35-10\\
 3.9& 1.52-12&4.11-12&5.54-12&1.03-11&1.93-11&1.02-9 &3.42-10&6.91-10&8.16-10\\
 4.0& 2.30-12&4.02-12&5.08-12&9.14-12&1.67-11&9.68-10&2.92-10&6.01-10&7.11-10\\
 4.1& 3.89-12&3.85-12&4.66-12&8.06-12&1.44-11&8.92-10&2.49-10&5.24-10&6.21-10\\
 4.2& 6.19-12&3.63-12&4.26-12&7.03-12&1.23-11&8.00-10&2.11-10&4.55-10&5.41-10\\
 4.3& 8.72-12&3.42-12&3.84-12&6.06-12&1.05-11&7.02-10&1.79-10&3.96-10&4.70-10\\
 4.4& 1.09-11&3.31-12&3.43-12&5.19-12&8.99-12&6.07-10&1.51-10&3.44-10&4.10-10\\
 4.5& 1.21-11&3.38-12&3.05-12&4.48-12&7.75-12&5.22-10&1.27-10&2.98-10&3.56-10\\
 4.6& 1.24-11&3.61-12&2.78-12&3.93-12&6.81-12&4.49-10&1.07-10&2.59-10&3.10-10\\
 4.7& 1.17-11&3.93-12&2.70-12&3.63-12&6.19-12&3.87-10&8.99-11&2.24-10&2.69-10\\
 4.8& 1.05-11&4.18-12&2.85-12&3.62-12&5.88-12&3.37-10&7.51-11&1.94-10&2.34-10\\
 4.9& 8.95-12&4.26-12&3.19-12&3.95-12&5.89-12&2.94-10&6.25-11&1.68-10&2.03-10\\
 5.0& 7.35-12&4.13-12&3.58-12&4.56-12&6.18-12&2.59-10&5.19-11&1.45-10&1.76-10\\
 5.1&-&-&-&-&-&2.30-10&4.29-11&1.25-10&1.52-10\\
 5.2&-&-&-&-&-&2.06-10&3.53-11&1.08-10&1.32-10\\
 5.3&-&-&-&-&-&1.84-10&2.89-11&9.32-11&1.14-10\\
 5.4&-&-&-&-&-&1.64-10&2.36-11&8.03-11&9.85-11\\
 5.5&-&-&-&-&-&1.43-10&1.92-11&6.91-11&8.51-11\\
 5.6&-&-&-&-&-&1.23-10&1.55-11&5.94-11&7.34-11\\
 5.7&-&-&-&-&-&1.03-10&1.25-11&5.10-11&6.33-11\\
 5.8&-&-&-&-&-&8.39-11&1.00-11&4.37-11&5.45-11\\
 5.9&-&-&-&-&-&6.72-11&7.96-12&3.74-11&4.69-11\\
 6.0&-&-&-&-&-&5.28-11&6.31-12&3.20-11&4.02-11\\
 6.1&-&-&-&-&-&4.09-11&4.99-12&2.73-11&3.46-11\\
 6.2&-&-&-&-&-&3.12-11&3.91-12&2.32-11&2.96-11\\
 6.3&-&-&-&-&-&2.36-11&3.06-12&1.98-11&2.53-11\\
 6.4&-&-&-&-&-&1.77-11&2.38-12&1.68-11&2.17-11\\
 6.5&-&-&-&-&-&1.32-11&1.84-12&1.42-11&1.85-11\\
 6.6&-&-&-&-&-&9.80-12&1.42-12&1.20-11&1.58-11\\
 6.7&-&-&-&-&-&7.23-12&1.09-12&1.02-11&1.34-11\\
 6.8&-&-&-&-&-&5.34-12&8.39-13&8.56-12&1.14-11\\
 6.9&-&-&-&-&-&3.94-12&6.39-13&7.20-12&9.64-12\\
 7.0&-&-&-&-&-&2.90-12&4.86-13&6.06-12&8.14-12\\
\hline
 \end{tabular}
\end{center}
 \end{table*}

\begin{table*}
\noindent{ Table 1 continues.}
\begin{center}
\scriptsize
\begin{tabular}{clllllllll}
\hline
$log_{10}T$ & \multicolumn{7}{c}{$\alpha_R(T)$} \\
& Fe I & Fe II & Fe III & Fe IV & Fe V & Fe XIII & Fe XXIV & Fe XXV & Fe XXVI \\
\hline
 7.1&-&-&-&-&-&2.04-12&3.69-13&5.13-12&6.88-12\\
 7.2&-&-&-&-&-&1.50-12&2.79-13&4.38-12&5.78-12\\
 7.3&-&-&-&-&-&1.09-12&2.10-13&3.79-12&4.85-12\\
 7.4&-&-&-&-&-&7.93-13&1.58-13&3.31-12&4.06-12\\
 7.5&-&-&-&-&-&5.79-13&1.19-13&2.89-12&3.38-12\\
 7.6&-&-&-&-&-&4.23-13&8.91-14&2.52-12&2.81-12\\
 7.7&-&-&-&-&-&3.08-13&6.67-14&2.17-12&2.32-12\\
 7.8&-&-&-&-&-&2.25-13&4.99-14&1.84-12&1.92-12\\
 7.9&-&-&-&-&-&1.65-13&3.72-14&1.53-12&1.57-12\\
 8.0&-&-&-&-&-&1.20-13&2.77-14&1.25-12&1.28-12\\
 8.1&-&-&-&-&-&8.81-14&2.05-14&1.01-12&1.04-12\\
 8.2&-&-&-&-&-&6.45-14&1.51-14&8.07-13&8.40-13\\
 8.3&-&-&-&-&-&4.74-14&1.11-14&6.36-13&6.73-13\\
 8.4&-&-&-&-&-&3.48-14&8.19-15&4.97-13&5.37-13\\
 8.5&-&-&-&-&-&2.57-14&6.04-15&3.85-13&4.26-13\\
 8.6&-&-&-&-&-&1.90-14&4.40-15&2.97-13&3.36-13\\
 8.7&-&-&-&-&-&1.41-14&3.22-15&2.27-13&2.63-13\\
 8.8&-&-&-&-&-&1.05-14&2.37-15&1.73-13&2.05-13\\
 8.9&-&-&-&-&-&7.86-15&1.72-15&1.32-13&1.59-13\\
 9.0&-&-&-&-&-&5.92-15&1.24-15&9.94-14&1.22-13\\ 
\hline
 \end{tabular}
\end{center}
 \end{table*}
 
\subsection{DR collision strengths}

 Recombination into high-\en bound levels of the \eion system 
is treated analytically using the
CC formulation that is an extension of electron impact excitation to
radiation damping of resonances and DR (BS 1985, Nahar and Pradhan 1994), 
as described in the Theory section.

 Fig. 7 shows a typical DR collision strength with high\en ($n > 
n_o \approx 10 $) resonances.
These calculations employ the same wavefunction expansion 
as the detailed photoionization/photorecombination calculations for all
possible low-\en (SLJ) levels with $n \leq n_o$.

\begin{figure} %
\psfig{figure=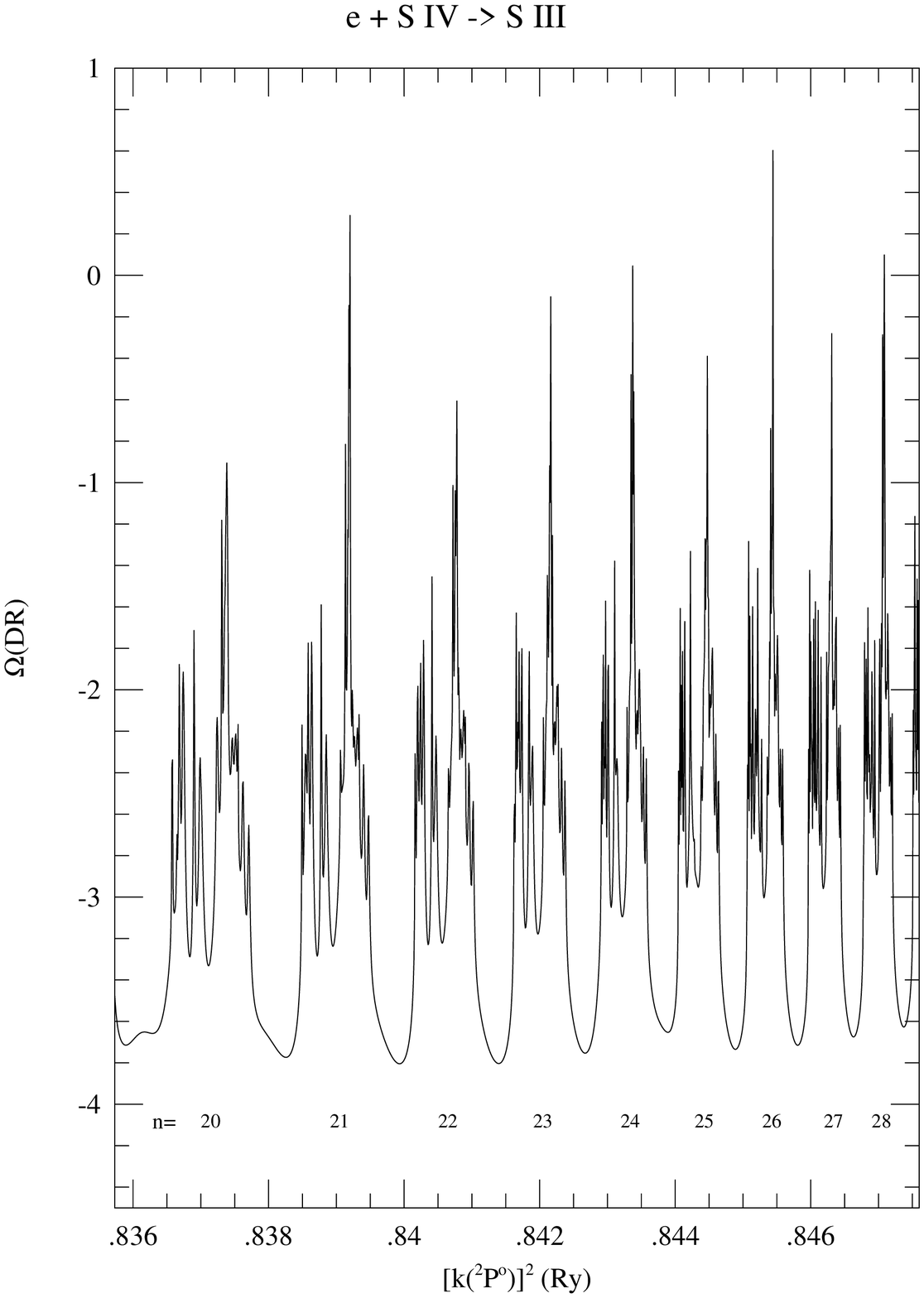,height=10cm,width=15.0cm}
\caption{DR collision strengths for high-\en ($10 < n \leq \infty $)
resonances in (e~+~S~IV) $\longrightarrow$ S~III recombination (Nahar
and Pradhan 1994).}
\end{figure}

\subsection{Photoexcitation-of-core resonances and DR}

 One of the main physical features of the present approach, self-consistent
treatment of photoionization and recombination, is exemplified by the
relationship between photoexcitation-of-core (PEC) resonances and the
inverse process of DR. Photoionization of bound levels along a Rybderg
series exhibit large PEC resonances at series limits corresponding to
dipole transitions in the core ion (Yu and Seaton 1987, Nahar and Pradhan
1992).  The PECs generally attenuate the cross section
by orders of magnitude relative to the background. 

\begin{figure} %
\psfig{figure=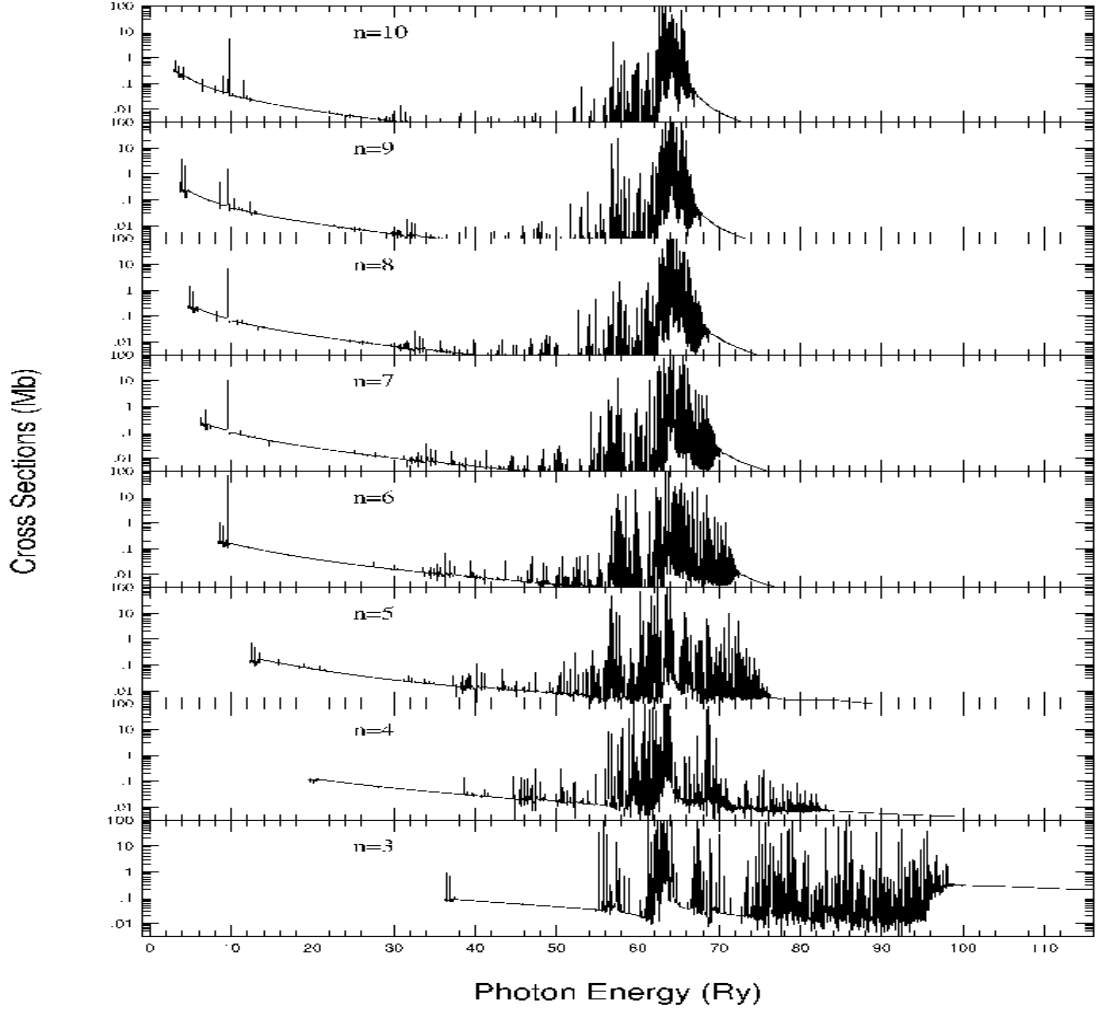,height=14cm,width=15.0cm}
\caption{Photoexcitation-of-core (PEC) resonances in photoionization of
the $2p^5 \ np \ (^3P_0)$ Rydberg series of levels of Fe~XVII. The giant
PEC resonance feature at approximately 63 Ry corresponds to strong
dipole excitations in the transition arrays $2p^5 - 2p^4 \ (3s,3d)$ within
the Fe~XVIII core (60CC results, Zhang \etal 2001).}
\end{figure}

 Fig. 8 illustrates
the PEC resonances in the photoionization of a Rydberg series of excited
bound levels of Fe~XVII.
The first noteworthy point is that the huge rise in the cross section
occurs in what is otherwise expected to be hydrogenic behavior, since
the photoionized levels belong to a Rydberg series for a given $nSL\pi$ 
or $nJ\pi$ symmetry of the \eion system; the higher the \en value
the more outstanding the PEC feature. This also points to the fact that
photoionization of excited levels of a non-hydrogenic ion may {\it not} be
treated in hydrogenic approximation, as is often done in practical
plasma applications.

 The inverse of PEC is DR. Electrons that excite the core ion may be
captured into high-\en autoionizing resonances, which subsequently decay
radiatively to corresponding high-\en bound levels of the \eion system. 
Such excitations generally involve strong dipole transitions in the
core and PEC resonances occur at photon energies equal to these
transitions.

\section{Physical effects and processes}

\subsection{The unified approach and separate RR and DR treatments}

The unified \eion recombination rate coefficients \art are valid over a 
wide range of temperatures for all practical purposes. 
In contrast, separate calculation of RR and DR rate coefficients are carried
out in different approximations valid for limited temperature ranges,
such as low-temperature DR, high-temperature DR, and RR. Moreover, 
division is sometimes made between $\Delta n = 0$ and $\Delta n \neq
0$ transitions in DR (Savin \etal 1999).
The unified cross sections generally cover all temperature
ranges, with cross sections computed over extended energy ranges. 

 However, the main problem
with separate treatment of RR and DR is more fundamental. Even if the DR
treatment is satisfactory,
the calculation of RR rate coefficient would require the calculation of
unphysical photoionization cross sections {\it without resonances}, 
computed in simpler approximations, such as the central field
method that does not include resonances, or a 
``one-channel" R-matrix calculation (Gorczyca
\etal 2002) that contradicts the CC
approach which, by definition, incorporates the necessary physics of
channel coupling as manifested in resonances.
A self-consistent and physical treatment to \eion recombination
is enabled by the use of CC wavefunctions employed in the unified
approach.

\subsection{Correspondence between photorecombination, dielectronic
recombination, and excitation}

The present close-coupling treatment of (electron-ion) recombination
is an unified and integrated approach to photorecombination (PR), DR, and
electron impact excitation (EIE). Fig.~9 illustrates the inter-relationships
required by conservation-of-flux and unitarity conditions for
PR, DR, and EIE for the (e~+~C~VI) $\rightarrow$ C~V
system (Zhang \etal 1999). The cross sections for all three 
processes, computed independently but with the same CC wavefunction expansion, 
are continuous functions of energy.
The PR cross sections include the background non-resonant contribution
as well as the resonances (left of the dashed line in Fig.~9), 
whereas the DR cross sections (right of the dashed line), 
computed using the BS
theory, neglect the background contribution. The two cross sections, the
PR and DR, match smoothly  at $\nu \approx 10.0$ showing that the background
contribution is negligible compared to the resonant contribution at high
$n > 10$. Further, the DR cross sections rise exactly up to the EIE cross
section at the threshold of excitation according to the theoretical
condition at threshold

\begin{equation}
 \lim_{n \rightarrow \infty} \Omega_{\rm DR}(n) = \lim_{k^2 \rightarrow 0}
\Omega_{\rm EIE}(k^2).
\end{equation}

The DR cross section in Fig.~9 at the
series limit $2\ ^1{\rm P}_1$ agrees precisely with the independently
determined value of the electron impact excitation cross section (filled
circle) for the dipole transition 
$1\ ^1{\rm S}_0 - 2\ ^1{\rm P}_1$, as required by the unitarity
condition for the generalized $S$-matrix and conservation of flux.
The continuous transition between the PR, DR, and EIE cross sections
serves to validate the accuracy of the BS theory of DR.
The DR cross sections
are, on the one hand, consistent with an extensively detailed coupled
channel treatment of photorecombination,
until an energy region where background recombination is
insignificant, and, on the other hand, 
consistent with the threshold behaviour at the EIE threshold. 
In fact Eq.~(27) provides a powerful accuracy check on the
possible importance of long range multipole potentials, 
partial wave summation, level degeneracies at threshold, 
and other numerical inaccuracies (discussed in previous works).

\begin{figure} %
\psfig{figure=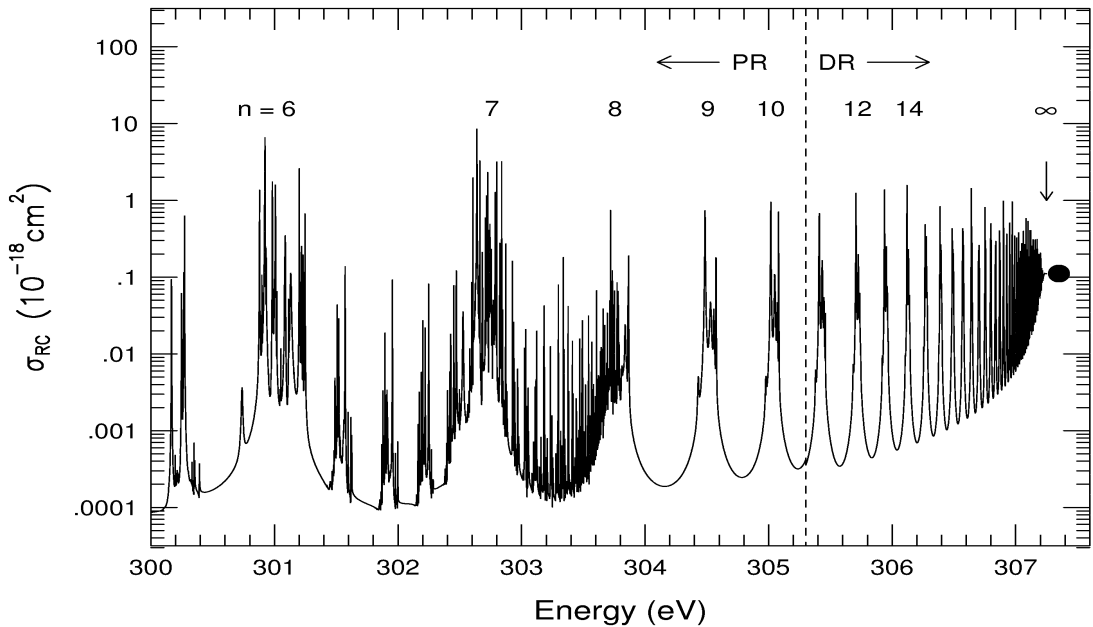,height=9cm,width=15.0cm}
\caption{Photo-recombination (PR), DR, and excitation cross sections
(Zhang \etal 1999), as
derived from photoionization calculations (left
of the dashed line), and the dielectronic (DR) cross sections
(right of the dashed line)
for (e + C~V) $\rightarrow$ C~IV; the filled circle represents
the near-threshold value of electron impact excitation cross section
for the dipole transition 1 $^1S_0 - 2 ^1P^o_1$ in C~V.}
\end{figure}

 The radiatively damped cross sections in Fig.~9
 illustrate that, owing to the interaction with the
radiation field, the autoionizing resonances are broadened, smeared, and
wiped out (in that order) as $n \rightarrow \infty$.
 At sufficiently high-$n$ the resonant contribution (DR) is very large
compared to the background, non-resonant photorecombination (PR)
cross section. In the unified method of electron-ion recombination,
for $ n > n_{\rm max}$, we employ the BS theory 
to compute the detailed and the averaged resonant DR
cross sections; the background contributions are computed in the hydrogenic approximation. 
The agreement and the continuity between the three
sets of data in Fig.~9 shows the CC method  for electron-ion
recombination to be accurate and self-consistent with 
inverse photoionization (photorecombination) and electron impact excitation.

 Another example that demonstrates the threshold behavior and continuity
of DR and electron impact excitation cross sections is the recent
calculation for (e~+~C~IV) $\longrightarrow$ C~III (Pradhan \etal
2001). In Fig.~10 we delineate the fine structure $\sigma_{DR}$ 
in the energy region
spanned by the fine structure $^2P^o_{1/2,3/2}$ thresholds. Fig. 10(a)
shows the detailed resonances in the vicinity of the two series limits.
Fig. 10(b) shows the $\sigma_{DR}$ averaged over the lower resonance series
$^2P^o_{1/2} n \ell$ below the $^2P^o_{1/2}$ level, but still with the
detailed resonance structures due to the higher series $^2P^o_{3/2} n
\ell$ (solid line). The $\sigma_{DR}$ averaged over both series is shown as the
dashed line. Above the $^2P^o_{1/2}$, $\sigma_{DR}$ is averaged over 
the $^2P^o_{3/2}
n \ell$ series. The sharp drop in the total $\sigma_{DR}$ at the $^2P^o_{1/2}$ 
threshold reflects
the termination of DR due to the $^2P^o_{1/2} n \ell$ resonance series, 
and with the $^2P^o_{3/2} n\ell$ contribution still low in spite of the
fact that $n \approx$ 96. The large drop in the DR cross section
is due to enhanced autoionization
into the excited level, when the $^2P^o_{1/2} n \ell$ channel opens up
at the lower fine structure threshold $^2P^o_{1/2}$ while the radiative 
decay remains constant.
 The $\sigma_{DR}(^2P^o_{3/2} n \ell)$  contribution builds
up to the second peak at $^2P^o_{3/2}$. 

In Fig.~10(b) it is shown that the resonance averaged 
$\lim_{n \rightarrow \infty}
<\sigma_{DR}(^2P^o_{1/2} n \ell)>$  = 242.57 Mb (dark circle at 
$^2P^o_{1/2}$), but the detailed $\sigma_{DR}$
has resonances due to the higher series $(^2P^o_{3/2} n \ell)$ lying
at and near threshold. The resonance averaged $\sigma_{DR}$ at the 
next DR peak, $\lim_{n \rightarrow \infty}
<\sigma_{DR}(^2P^o_{3/2} n \ell)>$  = 441.81 Mb (dark circle at
$^2P^o_{3/2}$). Interestingly, 
the fine structure in the theoretical $\sigma_{DR}$ in Fig. 10(a,b) 
appears to be discernible as a small dip in experimental data in 
just below 8 eV (Schippers \etal 2001, Pradhan \etal 2001).
Although the $^2P^o_{1/2,3/2}$ separation is only 0.013 eV, it may be
possible to detect these fine structure threshold effects in future 
experiments with increased resolution.

 At the 
$^2P^o_{1/2,3/2}$ thresholds the sum of the averaged fine structure
$<\sigma_{DR}>$ = $\sigma_{EIE}$ = 684.38 Mb.
Fig. 10(c) compares the near-threshold
EIE cross sections with the absolute measurements from two recent experiments,
(Greenwood \etal (1999) and Janzen \etal (1999)), 
convolved over their
respective beam widths of 0.175 eV and 2.3 eV.
Our results are in good agreement with both sets (and
also with another recent experiment by Bannister \etal (1998).
Although the present results are the first BPRM
calculations with relativistic fine structure for C~IV, 
their sum is in good agreement with previous LS coupling
CC calculations of $\sigma_{EIE}$ reported by (Burke 1992, Griffin \etal 2000,
Janzen \etal 1999).

\begin{figure} %
\psfig{figure=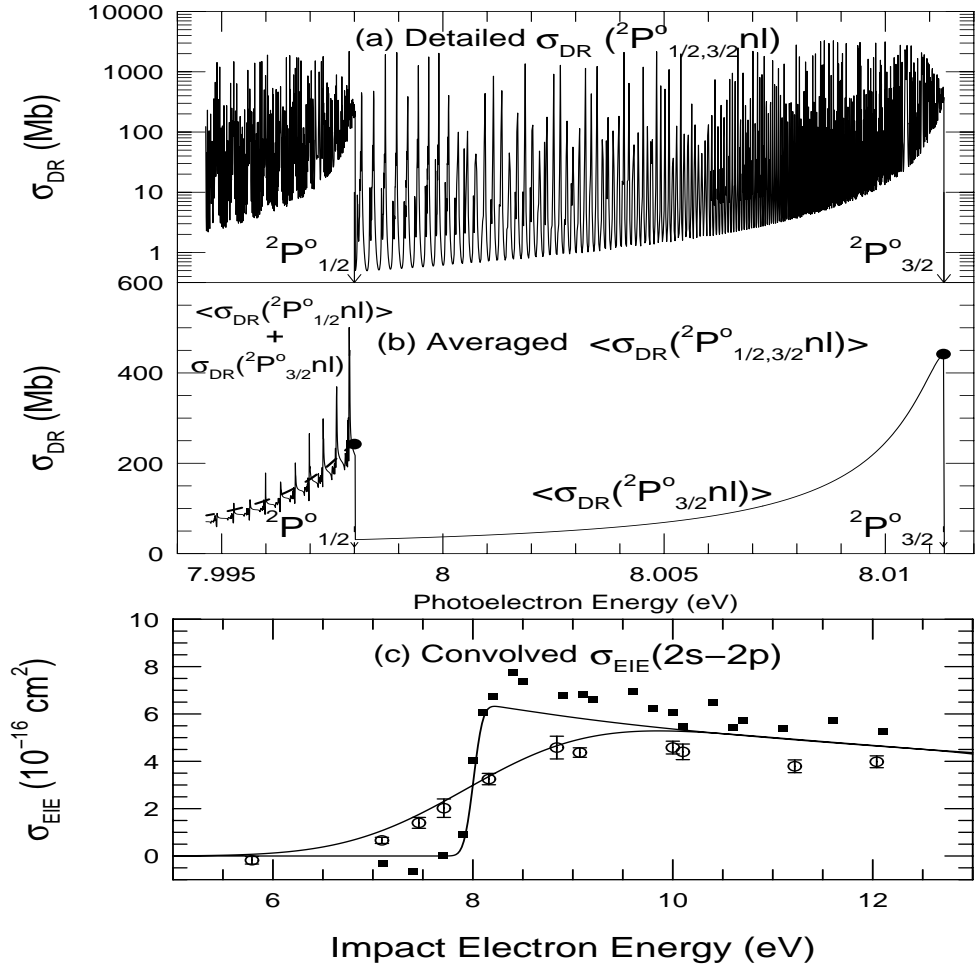,height=14cm,width=15.0cm}
\caption{DR and electron impact exciation cross sections $\sigma_{DR}$ and
$\sigma_(EIE)$ of C~IV: (a) detailed $\sigma_{DR}$ with
$^2P^o_{1/2,3/2} n\ell$ resonances; (b) $\sigma_{DR}$ averaged over
$^2P^o_{1/2} n\ell$ and detailed $^2P^o_{3/2} n\ell$ resonances (solid
line), average over the $^2P^o_{3/2} n\ell$ (dashed line); the dark circles
are the peak averaged $\sigma_{DR}$; (c) $\sigma_(EIE)$ convolved over 
experimental data with FWHM = 0.175 eV from Greenwood \etal (1999, 
filled squares), and with FWHM = 2.3 eV from Janzen \etal (1999,
open circles).}
\end{figure}

\subsection{Multiple DR bumps in \art}

 The general shape of the unified \eion recombination rate coefficient \art 
is illustrated in examples in the previous section. Basically, one finds
an exponential decrease with temperature starting with the 
background RR part of the rate coefficient, attennuated by a large bump at 
higher temperatures
due to DR. However, resonance series due to several ion thresholds might
contribute to \art over extended energy ranges. Nussbaumer and Storey
(1983) first pointed out that near-threshold low-energy resonances
give rise to a low-temperature DR bump. An extension of the same concept
is seen if several groups of resonances are interspersed throughout,
from low to high-energies. In such cases, we find multiple DR bumps in
\art. Fig.~11 shows (e~+~Fe~XXII) $\longrightarrow$ Fe~XXI
recombination, with discernible bumps due to the $n = 2$
and $n = 3$ groups of resonances (Nahar 2003).

\begin{figure} %
\psfig{figure=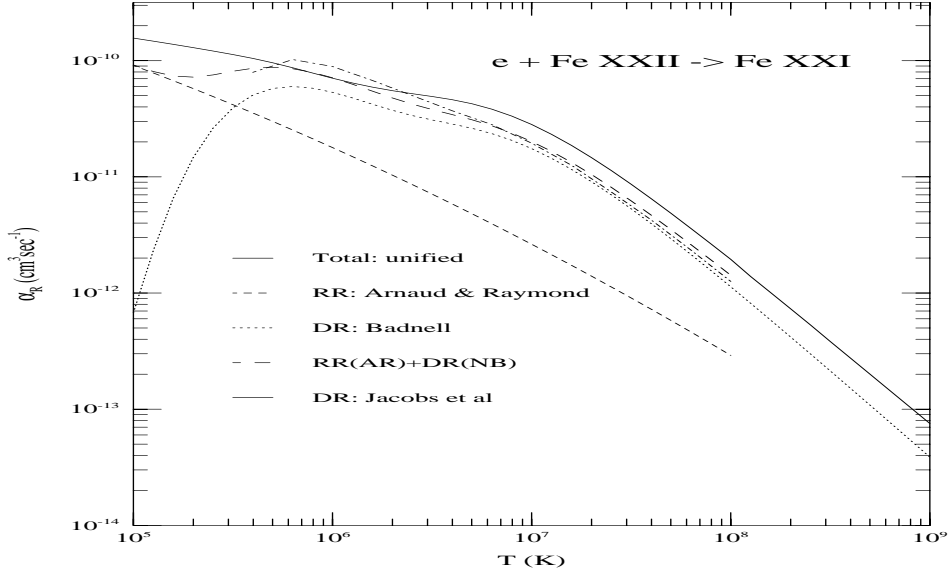,height=9cm,width=15.0cm}
\caption{Multiple DR bumps in the (e + Fe~XXII) $\rightarrow$ Fe~XXI 
recombination rate coefficients (Nahar 2003), 
compared with previous RR and DR values
(Arnaud and Raymond 1992, Badnell 1986, Jacobs \etal 1980).
The multiple bumps in \art are due to DR via
several groups of resonances in different energy regions in Fig.~12.}
\end{figure}

 The R-matrix eigenfunction CC expansion for B-like Fe~XXII includes 29 LS
terms (29-CC) up to the \en = 3 complex of configurations: $2s^2p, 2s2p^2, 2p^3,
2s2p3s, 2s2p3p, 2s2p3d, 2s^23s, 2s^3p, 2s^23d$. Including only the \en =
2 terms gives an 8-CC expansion. Fig.~12 shows the detailed photoionization
cross sections for two excited states, out of a total of 835 states of 
Fe~XXI calculated using the 29-CC Fe~XXII target expansion (the highest 8-CC 
threshold is also marked). It is particularly noteworthy how the
same resonance complexes appear quite differently in photoionization of 
states with different angular and spin symmetries SL$\pi$. 
It is clear that $\Delta n > 0$ transitions in the
core ion play an important role in the determination of total \art in 
Fig.~11 throughout the temperature range where Fe~XXI/XXI are likely to be
abundant in laboratory and astrophysical plasmas, and agree well with
the recently reported experimental values
in the "photoionized zone" T $\sim 30-80$ eV (Savin
\etal 2003, Fig. 5). At very low energies there are additional resonance
structures which may not be of practical importance but provide a good
check on theoretical works; more detailed calculations in this region 
($<$ 20 eV) are
in progress. However, at temperatures T $> 10^6$ K ($\sim 100$ eV) the
present rate coefficients in Fig. 11 are considerably higher than those 
reported by
Savin \etal (2003), since their measurements do not extend to high energies
relevant to collisionally ionized plasmas, particularly the
multiple DR bump region described herein; the temperature of maximum
abundance of Fe~XXI in collisional equilibrium is $\sim 10^7$ K (Arnaud
and Raymond 1992). A more detailed comparison will be reported elsewhere,
together with total \eion recombination rate coefficients for Fe~XXI at all 
temperatures of general importance (see also Section 4.6 on ionization
equilibrium).

\begin{figure} %
\psfig{figure=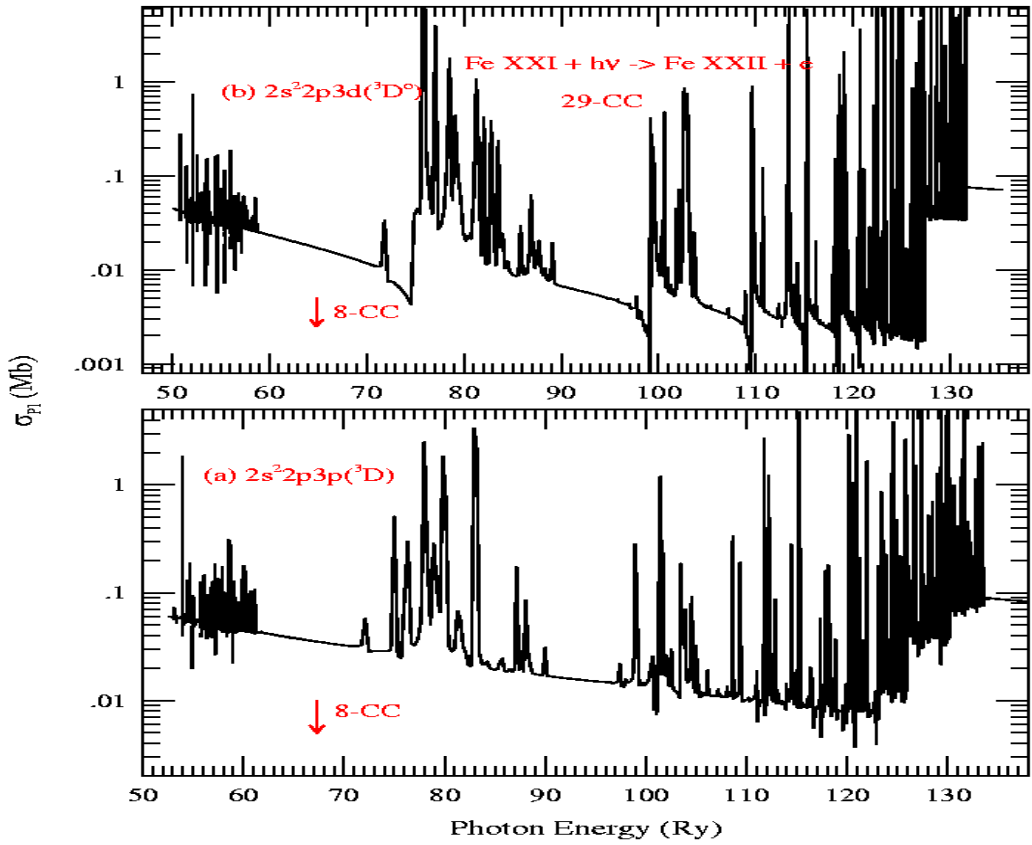,height=12cm,width=15.0cm}
\caption{Photoionization cross sections for two excited states of
carbon-like Fe~XXI, leading to multiple bumps in \art in Fig.~11 (Nahar 2003). 
The 8-CC refers to the \en = 2 thresholds, whereas the 29-CC refers to the
target states of Fe~XXII up to \en = 3 in the R-matrix CC expansion.}
\end{figure}

\subsection{Radiation damping of low-\en resonances}

Scattering, photoionization, and recombination of all atomic systems
involve an infinite number of resonances.
Close coupling calculations allow for resonances in an
{\it ab initio} manner. Broad low-lying
near-threshold resonances have autoionization rates $\gama \sim
10^{14} \ {\rm s}^{-1}$, much greater than typical dipole radiative 
decay rates
$\gamr \sim 10^{6-12} \  {\rm s}^{-1}$. For most atomic systems up to the
iron-peak
elements ($Z$ = 30), $\gamr < 10^{12} \ {\rm s}^{-1}$
for the lowest dipole transitions.
 Therefore, {\it a priori}, radiation damping
is not expected to be a dominant effect.
Notable exceptions are high-$Z$, H-like and He-like
ions with $\gamr \sim \gama$. For highly excited states 
with $n \rightarrow$ $\infty$, the $\gama$
decreases as $n^{-3}$ and $\gamr$ dominates.

 However, all resonances are susceptible to radiation damping somewhere
between 0 - 100\%, i.e. damping ratios (undamped/damped) from 1 to $\infty$,
depending on $n$, $\ell$,
energy relative to threshold, and radiative transition probability
$\gamr$ (radiation damping of resonance profiles was demonstrated in
Pradhan 1981, and Pradhan and Seaton 1985). However,
determination of damping factors of individual resonances per se may be
inaccurate as they depend on the precise positions, heights, and shapes of
resonances. 
Undamped/damped ratios $ > $ unity for a few individual resonance 
profiles do not imply that 
radiation damping is important in general. 
 More meaningful than the damping factors for individual resonances is
the combined effect of all damped resonances on the integrated rate
coefficient for an atomic process, i.e. on the quantities of
practical interest in laboratory and astrophysical applications.
 Rate coefficients are usually obtained by 
averaging cross sections over a Maxwellian. Generally broad near-threshold resonances
corresponding to the lowest $n$ and $\ell$ make
the dominant contribution, relative to narrower ones with higher $n$ and
$\ell$. The CC calculations include all closed channels
with $n \leq n_o$ and $\ell \leq n-1$.
Other methods based on quantum defect theory,
such as Gailitis averaging of
resonances, or the BS theory of DR, are employed for $ n_o < n \leq \infty$. 
As mentioned earlier, only the H-like or the
He-like core ions appear to require a detailed consideration of damping
of low-\en ($n < n_o$) resonances.

The primary
application of the present CC method for (e + ion) recombination is to
obtain unified total recombination rates for elements up to the
iron-peak elements with $Z \leq$ 30, well within the validity of the
intermediate coupling approximation using the BP method. 
 In general, the contribution from the near-threshold region 
 is dominated by large resonances with $\gama \sim 10^{14} \ {\rm s}^{-1}$;
 the extremely narrow 
resonances, with $\gama$ several orders of magnitude smaller, make relatively
little contribution and may be neglected. The calculations generally
resolve each resonance $n$-complex up to $\ell \leq 4$; the higher
$\ell$ resonances are
assumed to be damped out, thereby making allowance for radiation damping, 
although it is not a large effect on the final results. It is
interesting to note that experimental measurements also retain up to
about $\ell \leq 4$ levels before field ionization (Wolf \etal 1991).
With the exception of
H- and He-like ions  where  all resonances undergo significant radiative
damping, the cross sections and
rate coefficients should not be influenced by radiation damping when
the core radiative transitions do
not significantly compete with autoionization.  This is certainly true
for all $\Delta n = 0$ radiative core transitions even in highly charged
ions. For example, the theoretical BPRM
results without radiation damping of low-$n$ resonances in
photorecombination cross sections for (e~+~Ar~XIV) recombintion 
with a Boron-like core transition 2s - 2p (Zhang and Pradhan 1997),
are in excellent
agreement with the {\em absolute} cross sections measured from the heavy
ion storage ring CRYRING in Stockholm (DeWitt \etal 1995), 
both in magnitude and details of the extensive resonance structures
and background cross section.

\subsection{Resolution and analysis of resonance structures}

 The present unified approach incorporates the major effects of 
importance in the 
calculation of total effective recombination cross sections and rate 
coefficients. However, without loss of generality,
 some other physical effects might be of marginal interest in exceptional 
cases. 
 Criticism of the shortcomings of the method are confined to isolated
resonances, series limits, or small energy ranges, but which otherwise affect
the rate coefficients by no more than a few percent (Gorczyca
\etal 2002). For example, the precise choice of n$_o$ where the
background recombination cross section becomes negligible and DR
dominates for $n > n_o$, is not crucial and may be varied. Although some 
low-\en resonances close to n$_o$ might be very narrow and not fully
resolved, the effect is barely discernible (Zhang \etal 2001). 

 The precise
positions, shapes, and heights of resonances depend not simply on
resolution, but also on the CC wavefunction expansions (and accuracy thereof), 
relativistic fine structure, parameters chosen for $R$-matrix calculations, 
and numerical inaccuracies. 
Resolution {\it per se} is not a major problem by comparison (c.f.
Ramirez and Bautista 2002). 
The relevant radiative decay rates in neutrals, near-neutrals, and many-electron
systems are far smaller than the autoionization 
rates of near-threshold resonances.
For example, our detailed analysis for Fe~XVII shows that radiation damping 
is altogether unimportant. 

Finally, it might be pointed out that, as
yet, no {\it total} recombination rate coefficients of high accuracy
have been presented in
literature that are discrepant by about more than 10\% (the best
estimate of uncertainty in theory or experiments), due to radiation 
damping or other effects, with those calculated using the 
unified formulation. The narrow resonance structures may differ 
owing to resolution and accuracy of their energies, widths, heights, 
and shapes. The low-\en,$\ell$ resonances (\en $< n_o \approx 10$) are
usually highly resolved in the reported unified calculations and 
 adequately treated by the perturbative method employed.
Extremely narrow low-\en,$\ell$ resonances that might remain unresolved 
should be damped out, or have negligible effect.

\subsubsection{Plasma effects}

 While the unified method is quite general in scope, it does not
incorporate external field and density effects. These might be important
in practical situations, such as for DR near series limits. For example, 
in Fig.~3 the field-ionization cut-off in the experiment is estimated
at \en$_F \approx$ 19 (Schippers \etal 2001). The theoretical results
used the same approximate value for comparison (Pradhan \etal 2001).
However, a suitably general treatment of plasma fields and densities
for the calculation of total \eion recombination cross sections in the
present formulation is yet to be implemented.

\subsection{Ionization equilibrium}

 The present approach is especially suited to the calculation of
ionization fractions in astrophysical sources such as H~II
regions in general: diffuse and planetary nebulae, supernova remnants,
and broad line regions of active galactic nuclei.
   Ionization balance is the prime feature of astrophysical models of
these objects. The two most common assumptions for the
ionization conditions in the plasma are: (i) photoionization equilibrium, 
and (ii) collisional or coronal equilibrium
(Osterbrock 1989). The dominant ionizing process in the first case is
photoionization from the radiation field in the source, and in the second case, electron impact ionization in usually optically thin plasmas (such as
the solar corona). Both ionization processes are sought to be balanced by 
the inverse process of electron-ion recombination in the ambient medium
characterized by a given electron temperature, usually in terms of a Maxwellian
distribution, and an electron density. The two sets of ionization
balance equations may be written as

\begin{equation}
  \int_{\nu_0}^{\infty} \frac{4 \pi J_{\nu}}{h\nu} N(X^{z})
\sigma_{PI}(\nu,X^{z}) d\nu = \sum_j N_e N(X^{z+1}) \alpha_R(X_j^{z};T),
\end{equation}

and,

\begin{equation}
C_I(T,X^{z}) N_e N(X^{z}) = \sum_j N_e N(X^{z+1}) \alpha_R(X_j^{z};T),
\end{equation}

\noindent
where $\alpha_R(X_j^{z};T)$ is the total electron-ion recombination rate 
coefficient of the recombined ion of charge $z$, $X_j^{z}$, to state j 
at electron temperature T, $C_I$ is the rate coefficient for electron 
impact ionization, and $\sigma_{PI}$ is the photoionization cross section 
evaluated at photon frequency $\nu$ and convoluted with the isotropic 
radiation density J$_{\nu}$ of the source; N$_e$, $N(X^{z+1})$, and 
$N(X^{z})$ are the densities for the free electrons, and the recombining 
and recombined ions respectively. 

Implicit in these equations is the assumption that the ionization
rates on the left hand side refer to the ground state of the ion.
This condition is predicated on the assumption that the radiative and
collisional processes proceed on faster time scales than photoionization
and recombination. Substantial departures from these equilibrium
conditions may result at high densities where some excited states are
significantly populated or in LTE.  The sum on the
right hand side extends over the infinite number of excited states into
which recombinations can take place, depending on the temperature.

 Unlike previous works, the present work seeks to satisfy the 
photoionization equilibrium condition (Eq. 1)
in a fundamentally consistent manner: the photoionization and
recombination calculations are carried out  {\em using the same set of atomic
eigenfunctions}. Furthermore, as the detailed photoionization cross
sections include autoionizing resonances in an {\it ab initio}
manner, the electron-ion recombination rates subsume both the radiative
and the dielectronic recombination processes, that have been treated
separately in previous works using different methods.

In the coronal approximation (Eq. 2) some of the previous calculations
for ionization fractions are: Jacobs \etal (1980, and references therein), 
Shull and van Steenberg (1982a,b), Arnaud and Rothenflug (1985),
Sutherland and Dopita (1993). In general, most of the electron-ion 
recombination data employed in the earlier works is based on radiative 
recombination (RR) rates (e.g. Aldrovandi and Pequignot 1973) derived from 
photoionization cross sections that neglect the now well established 
phenomenon of autoionizing
resonances (e.g. Reilman and Manson 1979), and on dielectronic recombination 
(DR) rates derived from the Burgess general formula (Burgess 1965), but
incorporating the important advances made in the treatment of DR by Jacobs 
\etal, who showed the effect of autoionization into excited states, and later
by Nussbaumer and Storey (1983) who established the significant
contribution from near-threshold autoionizing resonances resulting in
a low-temperature bump (in addition to the usual high-temperature rise
in the DR rate).

 Using the present unified rate coefficients, ionization fractions for a
number of elements have been computed (e.g. Nahar and Pradhan 1997 for C and
N, Nahar 1999 for O). Fig.~13 shows a sample calculation in coronal
equilibrium (Nahar 1999). Ongoing calculations for astrophysically abundant
elemetns are being reported in a continuing seris of publications in the
Astrophysical Journal Supplements (e.g. Nahar and Pradhan 1997).

\begin{figure} %
\psfig{figure=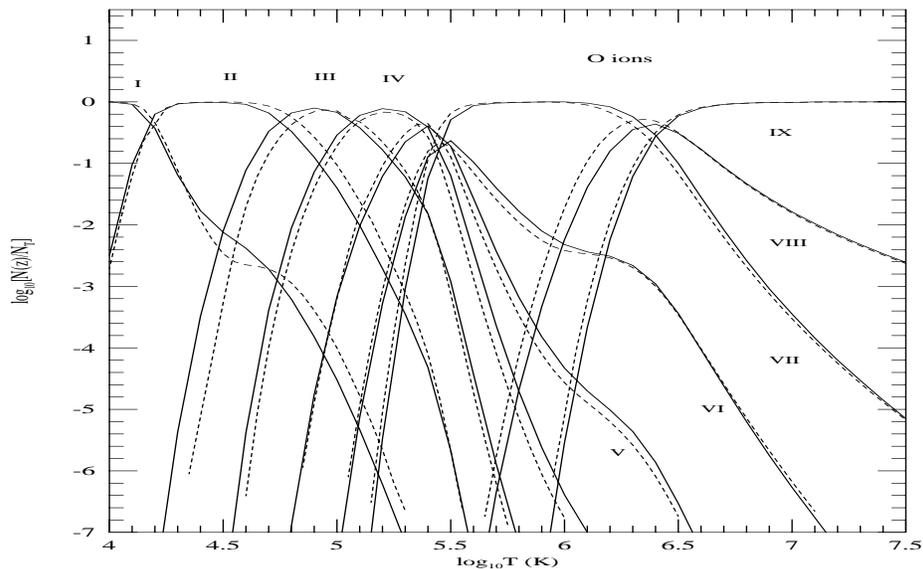,height=9cm,width=15.0cm}
\caption{Ionization fractions of oxygen ions in coronal or collisional
equilibrium (solid line, Nahar 1999) using unified \eion recombination rate
coefficients, compared to previous values (dashed line, Sutherland and
Dopita, 1993).}
\end{figure}

\section{Summary}

  The electron-ion recombination process is unified in nature; the
non-resonant and resonant contributions are 
inseparable and observed or measured together. However, these are
usually treated independently as
radiative and dielectronic recombination (RR and DR). A Unified
Approach, based on the Close-Coupling approximation and the R-matrix
method, has been developed and applied to the calculation of total
recombination cross sections and rates for over 50 atoms and ions.

 The ab initio calculations employ a coupled eigenfunction expansion for
detailed photoionization cross sections, with extensive delineation of
resonance structures, for a large number of bound levels of the (e+ion)
system, up to n(SLJ) = $n_o \approx$ 10. Detailed balance (Milne Relation)
thereupon yields photorecombination cross sections at all energies. 
For $n_o < n \leq \infty $, recombination is predominantly DR and is
treated using
the precise theory of Bell and Seaton; the non-resonant background 
contribution is treated hydrogenically as ``top-up" contribution.

 The advantages of the R-matrix method for (e+ion) recombination are:

\noindent 1. Unified treatment of non-resonant and resonant processes (RR and
DR) in an ab initio manner.\\
 2. Self-consistent treatment of photoionization and recombination
with identical wavefunction expansions.\\
 3. Relativistic fine structure effects are included using the 
Breit-Pauli R-matrix method.\\
 4. High-resolution of resonances to arbitrary accuracy, including
radiation damping when necessary (e.g. H- and He-like ions).\\
 5. Detailed agreement with experimental cross sections
    measured at synchrotron ion storage rings to 10-20\%.\\
 6. General validity for all ionization states of elements, from
neutrals to highly ionized.\\
 7. Total (e+ion) recombination rate coefficients are obtained at all 
temperatures for practical applications.\\
 8. Level-specific rate coefficients are
obtained for a large number of levels (typically few hundred) up to n(SLJ) 
= $n_o \approx 10$.\\
 9. Ionization fractions of elements in plasmas may be computed using 
consistent and accurate photoionization cross sections and total 
recombination rate coeffcients.\\
 10. Recombination spectra may be computed using level-specific reocmbination
rate coefficients; recombination-cascade matrices may be derived using 
transition probabilities that may also be computed with
the same R-matrix CC wavefunction expansion as 
photoionization, recombination, and excitation.

 The Unified method overcomes the shortcomings of simpler
methods based on independent resonance approximation that (i) divide
(e+ion) recombination into RR and DR, usually computed in different
approximations, (ii) further subdivision of DR into low-energy and
high-energy parts, such as $\Delta n$ = 0, and $\Delta n$ = 1, etc., and
(iii) neglect interference between non-resonant and resonant components,
likely to be of importance for neutrals and other ionization stages
with strong coupling effects, thereby limiting their validity to highly 
charged ions.

 The primary aim of this work is to obtain precise and complete total
recombination rates for practical applications.
 Following is the list of atoms and ions for which
self-consistent sets of $\sigma_{PI}$ and $\alpha_R(T)$ have so far
been obtained for over 50 ions (e.g. Nahar \& Pradhan 1997,
($http://www.astronomy.ohio-state.edu/\sim pradhan$):

\noindent
Carbon: C I, C II, C III, C IV, C V, C VI \\
Nitrogen: N I, N II, N II, N IV, N V, N VI, N VI \\
Oxygen: O I, O II, O III, O IV, O V, O VI, O VII, O VII \\
Si: Si I, Si II, Si IX \\
S: S II, S III, S XI \\
Ar: Ar V, Ar XIII
Ca: Ca VII, Ca XV
Fe: Fe I, Fe II, Fe III, Fe IV, Fe V, Fe XIII, Fe XVII, Fe XXI, Fe XXIV, 
Fe XXV, Fe XXVI \\
Ni: Ni II, Ni XXVI \\
C-like: F IV, Ne V, Na VI, Mg VII, Al VIII \\
Li-like: in progress

\noindent
The datasets for each ion include level-specific
unfiied recombination rate coefficients for typically hundreds of
bound levels with n$\leq$ 10.
 The self-consistent sets of photoionization/recombination datasets
also include
new photoionization cross sections that are generally an improvement
over the earlier Opacity Project data (The Opacity Project 1995,1996),
 since more extensive and accurate eigenfunction expansions are employed.
R-matrix transition probabilities are also available for many of the ions
listed.



\def\amp{{Adv. At. Molec. Phys.}\ }
\def\apj{{ Astrophys. J.}\ }
\def\apjs{{Astrophys. J. Suppl.}\ }
\def\apjl{{Astrophys. J. (Lett.)}\ }
\def\aj{{Astron. J.}\ }
\def\aa{{Astron. Astrophys.}\ }
\def\aasup{{Astron. Astrophys. Suppl.}\ }
\def\adndt{{At. Data Nucl. Data Tables}\ }
\def\cpc{{Comput. Phys. Commun.}\ }
\def\jqsrt{{J. Quant. Spectrosc. Radiat. Transf.}\ }
\def\jpb{{J. Phys. B}\ }
\def\pasp{{Pub. Astron. Soc. Pacific}\ }
\def\mn{{Mon. Not. R. Astron. Soc.}\ }
\def\pra{{Phys. Rev. A}\ }
\def\prl{{Phys. Rev. Lett.}\ }
\def\zpds{{Z. Phys. D Suppl.}\ }
\def\adndt{At. Data Nucl. Data Tables}


\end{document}